\documentclass[final]{lmcs}
\usepackage[utf8]{inputenc}

\usepackage{amsthm,amssymb}
\usepackage{ifdraft}
\usepackage{tikz}
\usepackage[labelfont+=small]{caption}
\usepackage{makecell}
\usepackage{mathtools}
\usepackage{booktabs}
\usepackage{etoolbox}
\usepackage{color}
\usepackage{xparse}

\ifdraft{
\usepackage[notref,notcite]{showkeys}
}{
\usepackage[final]{showkeys}
}

\usepackage{fixme}
\fxuselayouts{inline,nomargin}
\fxusetheme{color}
\FXRegisterAuthor{jr}{ajr}{JR}
\FXRegisterAuthor{tw}{atw}{TW}
\FXRegisterAuthor{rt}{art}{RT}
\FXRegisterAuthor{bk}{abk}{BK}
\ifoptionfinal{
  \fxsetup{silent=true}
}{}

\usepackage{hyperref}
\usepackage[capitalize,noabbrev,nameinlink]{cleveref}

\newboolean{spellingmode}
\setboolean{spellingmode}{false}

\ifthenelse{\boolean{spellingmode}}{%
	\nolinenumbers
	\AtEndPreamble{
	\hyphenpenalty=50000 %
	\pagestyle{empty}%
	\thispagestyle{empty}%
	\def\pagestyle#1{}%
	\def\thispagestyle#1{}%
	\def\labelmarginpar#1{}%
	}
}{}

\ifdraft{
\overfullrule=4pt %
}{
\overfullrule=0pt %
}

\usetikzlibrary{cd,automata,fit,backgrounds,calc}

\theoremstyle{defC}
\newtheorem{exaC}[thm]{Example}
\theoremstyle{thmC}
\newtheorem{obsC}[thm]{Observation}
\theoremstyle{definition}
\newtheorem{cons}[thm]{Construction}

\newcommand{\setrefname}[3]{\crefname{#1}{#2}{#3}\Crefname{#1}{#2}{#3}}
\setrefname{thm}{Theorem}{Theorems}
\setrefname{prop}{Proposition}{Propositions}
\setrefname{lem}{Lemma}{Lemmas}
\setrefname{cor}{Corollary}{Corollaries}
\setrefname{defi}{Definition}{Definitions}
\setrefname{exa}{Example}{Examples}
\setrefname{rem}{Remark}{Remarks}
\setrefname{asm}{Assumption}{Assumptions}
\setrefname{obs}{Observation}{Observations}
\setrefname{cons}{Construction}{Constructions}
\creflabelformat{enumi}{#2(#1)#3}

\crefalias{thmC}{thm}
\crefalias{propC}{prop}
\crefalias{lemC}{lem}
\crefalias{defiC}{definition}
\crefalias{exaC}{exa}
\crefalias{exas}{exa}
\crefalias{obsC}{obs}

\newcommand{\textqt}[1]{``#1''}
\newcommand{\takeout}[1]{}
\NewDocumentCommand{\ie}{s}{i.e.\IfBooleanTF{#1}{~}{\ }}
\NewDocumentCommand{\eg}{s}{e.g.\IfBooleanTF{#1}{~}{\ }}
\NewDocumentCommand{\cf}{s}{cf.\IfBooleanTF{#1}{~}{\ }}
\NewDocumentCommand{\wrt}{s}{w.r.t.\IfBooleanTF{#1}{\ }{~}}
\newcommand{\itemref}[2]{%
  \hyperref[#2]{\cref*{#1}\labelcref*{#2}}}

\newcommand{\fpair}[1]{\ensuremath{\langle #1 \rangle}}
\newcommand{\ceil}[1]{\ensuremath{\lceil #1 \rceil}}

\newcommand{\monoto}{\ensuremath{\rightarrowtail}}
\newcommand{\partialto}{\ensuremath{\rightharpoonup}}

\NewDocumentCommand{\set}{O{} m o}{%
  \ifthenelse{\equal{#2}{}}{%
    \ensuremath{{#1\emptyset}}%
  }{%
    \ensuremath{{#1\{#2\IfValueT{#3}{\mid #3}#1\}}}%
  }%
}

\newcommand{\C}{\ensuremath{\mathcal{C}}}
\newcommand{\D}{\ensuremath{\mathcal{D}}}
\newcommand{\E}{\ensuremath{\mathcal{E}}}
\renewcommand{\S}{\ensuremath{\mathcal{S}}}
\newcommand{\M}{\ensuremath{\mathcal{M}}}
\newcommand{\Bag}{\ensuremath{\mathcal{B}}}
\newcommand{\FBinTree}{H_B}
\newcommand{\FAut}[1]{H_{#1}}
\newcommand{\N}{\ensuremath{\mathbb{N}}}
\newcommand{\R}{\ensuremath{\mathbb{R}}}
\newcommand{\Pow}{\ensuremath{\mathcal{P}}}
\newcommand{\Set}{\ensuremath{\mathsf{Set}}}
\newcommand{\ar}{\ensuremath{\mathsf{ar}}}
\newcommand{\Path}{\ensuremath{\mathsf{Path}}}
\newcommand{\inj}{\ensuremath{\mathord{\mathsf{in}}}}
\newcommand{\pr}{\ensuremath{\mathord{\mathsf{pr}}}}

\newcommand{\inr}{\ensuremath{\mathord{\mathsf{inr}}}}
\newcommand{\inl}{\ensuremath{\mathord{\mathsf{inl}}}}
\NewDocumentCommand{\Coalg}{o}{\ensuremath{\mathsf{Coalg}}\IfValueT{#1}{_{#1}}}
\NewDocumentCommand{\Epi}{o}{\ensuremath{\mathsf{Epi}}\IfValueT{#1}{_{#1}}}
\NewDocumentCommand{\Mono}{o}{\ensuremath{\mathsf{Mono}}\IfValueT{#1}{_{#1}}}
\newcommand{\SplitEpi}{\ensuremath{\mathsf{SplitEpi}}}
\newcommand{\SplitMono}{\ensuremath{\mathsf{SplitMono}}}
\NewDocumentCommand{\Iso}{o}{\ensuremath{\mathsf{Iso}}\IfValueT{#1}{_{#1}}}
\NewDocumentCommand{\Mor}{o}{\ensuremath{\mathsf{Mor}}\IfValueT{#1}{_{#1}}}
\NewDocumentCommand{\Hom}{O{Hom} m m}{\mathsf{#1}\left(#2,#3\right)}
\renewcommand{\Im}{\ensuremath{\mathop{\mathit{Im}}}}

\newcommand{\id}{\ensuremath{\mathsf{id}}}
\newcommand{\Id}{\ensuremath{\mathsf{Id}}}

\NewDocumentCommand{\liftfactsys}{m m}{\ensuremath{#1\text{-carried}}}

\newsavebox{\mypullbackcorner}%
\sbox{\mypullbackcorner}{%
\begin{tikzpicture}
    \draw[-] (0,0) -- (.5em,.5em) -- (0,1em);
\end{tikzpicture}%
}
\newcommand{\pullbackangle}[2][]{\arrow[phantom,to path={
                     -- ($ (\tikztostart)!1cm!#2:([xshift=8cm]\tikztostart) $)
                        node[anchor=west,pos=0.0,rotate=#2,
                        inner xsep = 0]
                        {\begin{tikzpicture}[minimum
                        height=1mm,baseline=0,#1]
    \draw[-] (0,0) -- (.5em,.5em) -- (0,1em);
                        \end{tikzpicture}}}]{}}
\newcommand{\descto}[3][]{\arrow[phantom]{#2}[#1]{\text{\footnotesize{}#3}}}

\definecolor{coalgedge}{HTML}{4F677E}
\definecolor{coalgebraBackground}{HTML}{ededed}
\definecolor{lmcsblue}{HTML}{91B0CE}
\definecolor{lmcslightblue}{HTML}{B9CDE0}

\tikzset{
  loop at/.style={
    loop,
    out=#1-30,
    in=#1+30,
    looseness=5,
    every node/.append style={
      anchor=#1-180,
    },
  },
}

\tikzstyle{shiftarr}=[
        rounded corners,%
        to path={--([#1]\tikztostart.center)
                     -- ([#1]\tikztotarget.center) \tikztonodes
                     -- (\tikztotarget)},
]

\tikzset{
      vertex/.style={
        fill=black,
        shape=circle,
        outer sep = 1mm,
        inner sep = 0mm,
        minimum size=2mm,
      },
      named vertex/.style={
        inner sep = 0mm,
        outer sep = 1mm,
      },
      set/.style={
        draw=black!20,
        fill=none,
        line width=1mm,
        rounded corners=2mm,
        inner sep = 2mm,
      },
      myedge/.style={
        ->,
        draw=black,
        every node/.append style={
          shape=circle,
          inner sep=1pt,
          fill=black,
          text=white,
          draw=none,
          sloped,
          minimum width=5mm,
        },
      },
}

\tikzset{
  scope of math nodes/.style={
    every node/.append style={
      execute at begin node=$,%
      execute at end node=$%
    },
  },
  coalgebra/.style={
    block line/.style={
      draw=black!50,
      line width=1.2pt,
    },
    block/.style={
      block line,
      rounded corners=3pt,
      inner sep=1pt,
      minimum height=6mm,
      minimum width=6mm,
    },
    scissors line/.style={
      draw=black!50,
      text=black!50,
      font=\footnotesize,
      line width=0.8pt,
      shorten <= -4pt,
      shorten >= -4pt,
      dotted,
    },
    state/.style={
      outer sep=2pt,
      inner sep=2pt,
      minimum size=4mm,
      shape=circle,
      draw=coalgedge!70,
      fill=white,
      line width=1pt,
      text=black,
    },
    add state borders/.style={
      state/.append style={
        draw=blue,
        shape=circle,
        outer sep=3pt,
        line width=.4pt,
        inner sep=2.4pt, %
      },
    },
    final/.style={
      draw=coalgstate,
      double=white,
      inner sep=2pt,
      double distance=1pt,
      line width=.4pt,
    },
    transition/.style={
      ->,
      >={Triangle[scale=0.7]},
      line width=0.8pt,
      draw=black,
      every node/.append style={
        on transition/.style={
          shape=circle,
          inner sep=.5pt,
          minimum width=2mm,
          minimum height=2mm,
          draw=none,
          fill=black,
          text=white,
          font=\sffamily,
        },
      },
    },
    transition predraw/.style={
      preaction = {
        draw,
        -,
        draw=coalgebraBackground,line width=4.6pt,
        line cap=round,shorten <= 2mm, shorten >= 2mm,
        },
    },
    path with edges/.style={
      every edge/.append style={transition}
    },
  },
  initial by arrow/.append style={
    initial text={},
    initial distance=3mm,
  },
  every initial by arrow/.append style={
    transition,
    every node/.append style={overlay}, %
    shorten <= 0pt,
  },
  coalgebra frame/.style={
    fill=coalgebraBackground,
    rounded corners=4pt,
    line width=1pt,
  },
  coalgname/.style={
    outer sep=4pt,
    inner sep=0pt,
    rounded corners=1pt,
    anchor=north west,
    text=black,
  },
}

\tikzset{
  class diagram/.style={
    concept/.style={
      align=center,
      fill=coalgebraBackground,
      rounded corners=4pt,
      line width=1pt,
      minimum width=.25\textwidth,
    },
  },
  realm/.style={
    draw=coalgedge,
    line width=1.5pt,
    rounded corners=4pt,
  },
  realm title/.style={
    text=coalgedge,
    anchor=south,
    rotate=90,
    font=\sffamily\bfseries,
  },
  generalizes edge/.style={
    >={Triangle[open,length=3mm,width=3mm]},
    ->,
    line width=1pt,
    draw=lmcsblue,
    shorten >= 1mm,
    shorten <= 1mm,
    preaction={
      -,
      draw=white,
      line width=4pt,
      shorten >= 2mm,
      shorten <= 2mm,
    },
  },
}

\newsavebox{\generalizesto}
\sbox{\generalizesto}{%
    \begin{tikzpicture}[class diagram]
      \path[generalizes edge,shorten <=0,shorten >=0] (0,0) -- +(8mm,0); 
    \end{tikzpicture}
}

\begin{document}

\title[From Minimality and Reachability to Trees in Coalgebra]
      {From Minimality and Reachability\texorpdfstring{\\}{} to Trees in Coalgebra}
\keywords{trees, coalgebra, tree unravelling} %
\thanks{Supported by the NWO grants VI.Vidi.223.096 and OCENW.M20.053.} %

\author[T.~Wi{\ss}mann]{Thorsten Wi{\ss}mann\lmcsorcid{0000-0001-8993-6486}}[a]
\author[B.~Kocsis]{B{\'a}lint Kocsis\lmcsorcid{0000-0003-0570-3031}}[b]
\author[J.~Rot]{Jurriaan Rot\lmcsorcid{0000-0002-1404-6232}}[b]
\author[R.~Turkenburg]{Ruben Turkenburg\lmcsorcid{0000-0001-7336-9405}}[b]

\address{Friedrich-Alexander-Universit{\"a}t Erlangen-N{\"u}rnberg, Germany}
\email{uni@thorsten-wissmann.de}
\urladdr{https://thorsten-wissmann.de}
\address{Radboud Universiteit, Nijmegen, The Netherlands}
\email{mail@balintkocsis.site, jurriaan.rot@ru.nl, ruben.turkenburg@ru.nl}
\urladdr{https://balintkocsis.site, https://jurriaan.me, https://rubenturkenburg.info}

\begin{abstract}
\bknote{}%
  An automaton is called \emph{reachable} if every state is reachable from the initial state. This notion has been generalized coalgebraically in two ways: first, via a universal property on pointed coalgebras, namely, that a reachable coalgebra has no proper subcoalgebras; and second, a coalgebra is reachable if it arises as the union of an iterative computation of successor states, starting from the initial state.

  In the current paper, we present corresponding universal properties and iterative constructions for trees. The universal property captures when a coalgebra is a tree, namely, when it has no proper tree unravellings. The iterative construction unravels an arbitrary coalgebra to a tree. We show that this yields the expected notion of tree for a variety of standard examples.

  We obtain our characterization of trees by first generalizing the previous
  theory of reachable coalgebras and of a minimal object in a category, related
  to projectivity.
  Surprisingly, both the universal property and the iterative construction for
  trees arise as instances of this generalized notion of reachability.
  Our iterative construction works for all analytic set functors.
\bknote{}%
\end{abstract}

\maketitle

\section{Introduction}
\label{sec:intro}
\bknote{}%
\emph{Reachability} and having a \emph{tree structure} are two fundamental structural properties of state-based systems.
For instance, a deterministic automaton is reachable if for
every state, there exists a path from the initial state to it, and it is a tree if for every state, there is a \emph{unique}
path from the initial state (the latter makes sense in particular if the transition map is partial, otherwise it forces the automaton to be infinite).
Analogous notions of reachability and tree structure can be defined for many variations of
automata and graphs.

Reachability has been characterized at the level of pointed coalgebras -- intuitively, coalgebras with an initial state -- providing a general notion that instantiates to a wide variety of systems beyond deterministic automata.
There are, in fact, two equivalent characterizations (\cf \cref{fig:table}). First, a pointed coalgebra is
reachable if it has no proper pointed subcoalgebras~\cite{AMMS03}. Second, a pointed coalgebra is reachable if, intuitively, the entire state space is obtained when starting at the initial state and iteratively computing next states~\cite{wmkd20reachability,BarloccoEA19}. The latter construction makes use of the notion of \emph{least bounds} to compute one-step successor
states (originally introduced under the name \emph{base} by Blok~\cite{Blok12}). While the first approach
characterizes reachability abstractly via a universal property, the second approach provides an iterative process that can be used to obtain the reachable part of a pointed coalgebra.

In the current paper, we extend this theory by providing characterizations of when the transition structure of a coalgebra is tree-shaped.
Analogously to the case of reachability,
we provide two equivalent characterizations, one via a universal property\bknote{} and one via an iterative construction. These are summarized
in the second row of \cref{fig:table}.

\begin{table}\centering%
  \begin{tabular}{@{}lccc@{}}%
    \toprule
    Notion
    & Universal Property
    & Morphism Class
    & Iterative Construction
    \\
    \midrule
    Reachable
    & \makecell[c]{No proper subcoalgebra\\\cite{AMMS03}}
    & Monic
    & \makecell[c]{Union of least bounds \\ \cite{BarloccoEA19,wmkd20reachability}}
    \\
    Trees
    & No proper unravelling
    & Arbitrary
    & Coproduct of $F$-precise maps
    \\
    \bottomrule
  \end{tabular}
  \caption{Properties of coalgebras and their constructions.}
  \label{fig:table}
\end{table}

The first characterization (\cref{defTree}) intuitively says that a pointed coalgebra is a tree if 
it has no proper unravellings, where an unravelling can copy shared nodes and unfold loops.
Formally, an unravelling of a coalgebra is captured as a pointed coalgebra homomorphism into it.
The difference with the characterization of reachability is that this homomorphism is not required
to be monic.
The second characterization (\cref{corTreeCoprodLevels}) is again an iterative construction: a coalgebra is a tree if it
arises as the \emph{coproduct} of its iteratively computed successors. There is, however, a catch: the classic least bound does not
account for avoiding sharing of one-step successors. Therefore, we use the notion of \emph{precise maps}~\cite{WDKH2019},
a variation of least bounds whose relevance for tree unravelling has been conjectured~\cite[Ex.~4.16,~Fn~2]{Wissmann22}.
The resulting iterative construction (\cref{constrTree}) can be used to compute the tree unravelling of a given pointed coalgebra.

We recover expected tree notions for instances such as (partial) deterministic automata, coalgebras for various simple polynomial functors, and coalgebras for the bag functor.
Interestingly, coalgebras for the powerset functor are (almost) never a tree in our characterization, as this functor allows a form of copying that yields multiple edges between two nodes.

The key idea behind our technical approach is that all characterizations, of reachability and of being a tree,
turn out to be instances of a generalized notion of reachability (\cref{defReachable}).
While reachability is studied in~\cite{wmkd20reachability} in the setting of an $(\E,\M)$-factorization system where $\M$ consists of monos (not necessarily including all of them),
here we relax this condition to allow $\M$ to be arbitrary. Then, taking $\M$ to be \emph{all} morphisms is precisely what yields the characterization of trees (see also the morphism class column in \cref{fig:table}).
In order to make the generalization work, we have to relax the condition that every $\M$-subcoalgebra be an isomorphism to the requirement that it be a split epimorphism.
When $\M$ consists of monos, this is equivalent to the isomorphism condition, recovering the classical notion of reachability.

As reachability of coalgebras has previously been related to the notion of
\emph{minimality in a category}~\cite{Wissmann22} (as indicated in \cref{fig:classDiagram}), the generalization of reachability
also leads to a generalized notion of minimality (\cref{defMinimal}).
Even though our primary focus is coalgebras, we investigate some properties of
this new notion at a higher level of abstraction. One of the (perhaps surprising)
results originating from this approach is the connection between minimality and the
notion of \emph{generalized projectivity} with respect to a class of morphisms
(\cref{thm:projective}).

We then show (\cref{levelConstruction}) that the iterative construction can also be formulated at this level, and that it is equivalent to the characterization via a universal property (\cref{corReach}),
generalizing the corresponding results from~\cite{wmkd20reachability,BarloccoEA19}. We thus arrive at a general theory that instantiates both
to reachability and to the property of being a tree, and provides two equivalent characterizations, one via a universal property and one via an iterative construction.

\bknote{}\bknote{}%

\begin{figure}
  \begin{tikzpicture}[class diagram,x=.33\textwidth,y=3cm]
  \node[concept] (simple) at (0,0) {
    Simple\\
    Coalgebra\\
    \cite[p.~34]{Gumm03}
  };
  \node[concept] (reachable) at (1,0) {
    Reachable\\
    Coalgebra\\
    \cite[p.~26]{AMMS03}
  };
  \node[concept] (tree-shaped) at (2,0) {
    Tree-Shaped\\
    Coalgebra\\
    \textit{Previous Version}
  };
  \node[concept] (confer) at (1.666,-0.94) {
    $\M$-Minimal \\
    Coalgebra \\
    \textit{Previous Version}
  };
  \node[concept] (min) at (0.33,-2) {
    (Iso-)Minimal \\
    Object \\
    \cite[Def.\ 4.1]{Wissmann22}
  };
  \node[concept] (this) at (1.666,-2) {
    $\M$-Minimality
    \\
    (\cref{defMinimal})
    \\
    \textit{Present Extended Paper}
  };
  \node[realm,
        fit={(simple) (tree-shaped) ([yshift=-6mm]confer.south)}
        ]
    (coalg) {};
  \node[realm title] at (coalg.west) {Coalgebra};
  \node[realm,
        fit={(min) (this) (min.west -| simple.west) (tree-shaped.east |- min)
            ([yshift=6mm]min.north)},
       ] (objects) {};
  \node[realm title] at (objects.west) {Objects};
  \path[generalizes edge] (this) -- (min);
  \path[generalizes edge] (this) -- (confer);
  \path[generalizes edge] (min) -- (simple);
  \path[generalizes edge] (min) -- (reachable);
  \path[generalizes edge] (confer) -- (reachable);
  \path[generalizes edge] (confer) -- (tree-shaped);
  \end{tikzpicture}
  \caption{Relations between different minimality concepts in the realm of coalgebra and objects in an abstract category. An edge
    $X\usebox{\generalizesto}Y$ indicates that concept $X$ generalizes concept
    $Y$. \emph{Previous Version} refers to the conference paper~\cite{WKRT25}
    that the present paper is extending.}
  \label{fig:classDiagram}
\end{figure}

\paragraph*{Related Work}
Our approach to trees seems orthogonal to \emph{arboreal
categories}\ \cite{DBLP:conf/icalp/AbramskyR21}, which use \emph{lax} coalgebra morphisms. In this paper, the properties of trees and tree unravelling come from the
properties of \emph{strict} coalgebra morphisms.
\bknote{}%

The present article is an extended version of a conference paper~\cite{WKRT25}.
In addition the the inclusion of all proofs, the present paper contains new material:
\begin{enumerate}
\item We further generalize the definition of
reachability from coalgebras to a general \emph{minimality} notion on
objects of an abstract category (\cref{fig:classDiagram}).
\item We characterize minimality via projectivity (\cref{thm:projective})
\item We show the connection between precise factorizations and
wide pullbacks (\cref{thmPreciseIffWidePullback}). In particular, we prove that a finitary set-functor admits precise factorizations iff it is analytic (\cref{setPreciseFactor}).
\end{enumerate}

\bknote{}\jrnote{}

\paragraph*{Structure of the paper}
In \cref{sec:prelims}, we recall preliminaries on coalgebras and factorization systems.
In \cref{sec:precise}, we extend the theory of precise morphisms and relate it
to the notion of least bounds and analytic functors.
In \cref{sec:genmin}, a generalized notion of \emph{minimality in a category}
is introduced, which encompasses both reachability and previous notions of minimality.
In \cref{sec:reach}, we define generalized reachability by instantiating minimality to categories of coalgebras,
and we complement the theory with a matching iterative construction.
We further instantiate this generalized notion of reachability in \cref{sec:trees} to obtain an account
of trees in coalgebras. The paper concludes with a description of future work in \cref{sec:fw}.
\bknote{}\bknote{}%

\section{Preliminaries}\label{sec:prelims}
In the following, we assume basic knowledge of category theory~(\cf standard
textbooks such as~\cite{joyofcats,awodey2010category}).
Usually, $\C$ and $\D$ denote categories and $F$ denotes a functor.

\begin{nota}
  Product projections are denoted by $\pr_i\colon \prod_{j} A_j \to A_i$.
  Coproduct injections are called $\inl\colon A\to A+B$, $\inr\colon B\to A+B$, and $\inj_i\colon A_i\to \coprod_{j} A_j$.
  We write $\ceil{n}$ for the set $\set{0,\dotsc, n-1}$.
  Finally, the classes of all epimorphisms, monomorphisms,
  split epimorphisms, split monomorphisms, isomorphisms,
  and arbitrary morphisms in a category are denoted by
  $\Epi$, $\Mono$, $\SplitEpi$, $\SplitMono$, $\Iso$, and $\Mor$, respectively.
\end{nota}

\begin{defi}\label{defCoalg}
  Given a category $\C$ and an endofunctor $F \colon \C \to \C$,
  an \emph{$F$-coalgebra} is a pair $(C,c)$ consisting of an object $C$ (of $\C$) and a morphism $c\colon C\to FC$
  (in $\C$). For a fixed object $I\in \C$, a \emph{pointed $F$-coalgebra} is an $F$-coalgebra equipped with a morphism
  $i_C\colon I\to C$, called the \emph{point}.
  An \emph{$F$-coalgebra morphism} $h\colon (C,c)\to (D,d)$ between
  $F$-coalgebras $(C,c)$ and $(D,d)$ is a morphism $h\colon C\to D$ with $d\cdot
  h = Fh\cdot c$:
  \[
    \begin{tikzcd}
      C \arrow{d}[swap]{h}\arrow{r}{c}& FC \arrow{d}[overlay]{Fh}\\
      D \arrow{r}[swap]{d} & FD
    \end{tikzcd}%
  \]
  A \emph{pointed $F$-coalgebra morphism} $h\colon (C,c,i_C)\to (D,d,i_D)$ is an $F$-coalgebra
  morphism $h \colon (C,c) \to (D,d)$ that preserves the point, \ie such that $i_D = h\cdot i_C$:
  \[
    \begin{tikzcd}[row sep=1.9pt] %
      &C \arrow{dd}{h} \\
      I
      \arrow{ur}{i_C}
      \arrow{dr}[swap]{i_D}
      \\
      &D
    \end{tikzcd}%
  \]
  We denote the category of $F$-coalgebras and $F$-coalgebra morphisms
  by $\Coalg(F)$, and the category of pointed $F$-coalgebras and pointed $F$-coalgebra morphisms by $\Coalg[I](F)$.
\end{defi}
The functor $F$ is dropped from the terminology when understood from context.
Intuitively, the \emph{carrier} $C$ of a coalgebra $(C,c)$ is the state space
and the point $I\to C$ models an initial state. For $\C=\Set$, we fix $I
= 1$ in this paper. Alternatively, one could choose \eg $I = 2$ to model two initial states.
The morphism $c\colon C\to FC$ sends states to
their possible next states. The functor of choice $F$ defines how these possible
next states $FC$ are structured. We provide a few examples below, and refer to~\cite{DBLP:books/cu/J2016,DBLP:journals/tcs/Rutten00} for more examples
and a general introduction to the theory of coalgebras.

\begin{exas}\label{ex:coalgebras}\hfill
  \begin{enumerate}
  \item The powerset functor $\Pow\colon \Set\to\Set$ sends each set to its
  powerset. A $\Pow$-coalgebra $c \colon C \rightarrow \Pow C$ is a transition system, where $C$ is the set of states,
  and there is a transition from~$x$ to~$y$ iff $y \in c(x)$. It can also be viewed as a directed graph (where self-loops are allowed).

  \item\label{ex:bintree} Consider the functor $\FBinTree \colon \Set \to \Set$ given by $\FBinTree X = X \times X + \{\bot\}$.
  An $\FBinTree$-coalgebra $c \colon C \rightarrow C \times C + \{\bot\}$
  maps every state to either a pair of next states, or to $\bot$, representing termination.
  
  \item\label{ex:pdfa} Let $A$ be a fixed set (the alphabet), let $2 = \{0,1\}$, and let $\FAut{A} \colon \Set \to \Set$ be given by $\FAut{A}X = 2 \times (X + \{\bot\})^A$.
  Coalgebras for~$\FAut{A}$ are \emph{partial deterministic automata} over~$A$. They are conveniently
  presented as pairs of maps $\fpair{o, \delta}\colon C \rightarrow 2 \times (C + \{\bot\})^A$. Here, $C$ is the set of states,
  and $o \colon C \rightarrow 2$ is thought of as the output function; we say $x \in C$ is an accepting state
  iff $o(x)=1$. For~$x \in C$ and a letter~$a \in A$, if $\delta(x)(a) = \inl(y)$ for some~$y \in C$, then we say that there
  is an $a$-transition from~$x$ to~$y$. If $\delta(x)(a) = \inr(\bot)$, then the $a$-transition of~$x$ is undefined.
  A point $q_0\colon 1\to C$ models the initial state.
  
  We extend $\delta$ to a function $\delta^*\colon C\to (C+\set{\bot})^{A^*}$ on words in the usual way:
  \begin{align*}
    \delta^*(q)(\varepsilon) &= \inl(q), \\
    \delta^*(q)(aw) &=
      \begin{cases}
        \delta^*(q')(w) & \text{ if } \delta(q)(a) = \inl(q'), \\
        \inr(\bot) & \text{ otherwise.}
      \end{cases}
  \end{align*}

  \item A \emph{signature} is a set~$\Sigma$ together with a map $\ar\colon
  \Sigma\to \N$, sending each symbol~$\sigma \in \Sigma$ to its
  arity~$\ar(\sigma)\in \N$. Every signature~$\Sigma$ induces a polynomial
  $\Set$-functor~$F_\Sigma$ given by $F_\Sigma X = \coprod_{\sigma\in \Sigma} X^{\ar(\sigma)}$.
  Pointed $F_\Sigma$-coalgebras can be understood as automata
  for which a state~$q$ marked by a symbol~$\sigma\in \Sigma$ offers
  $\ar(\sigma)$-many inputs.
  For a finite input alphabet~$A$ and the signature $\Sigma = 2$, $\ar(b) = |A|$,
  the pointed $F_\Sigma$-coalgebras are precisely the deterministic automata over~$A$, because every state is marked with either $0\in 2$ (\textqt{non-accepting}) or $1\in 2$ (\textqt{accepting}) and has transitions for the $|A|$ different inputs.
  Partial deterministic automata for~$A$ arise as
  $F_\Sigma$-coalgebras for the signature $\Sigma = 2\times \Pow A$ with
  $\ar(b,S) = |S|$.
  \end{enumerate}
\end{exas}

\begin{defi}\label{defMultigraph}
  A \emph{directed multigraph}~$G = (V,E,s,t)$ consists of a set~$V$
  of \emph{vertices}, a set~$E$ of \emph{edges}, and functions $s,t\colon E\to V$
  sending each edge to its source and target vertex, respectively. A pointed (or rooted)
  graph additionally has a distinguished vertex~$v_0\in V$.
  We call a graph \emph{locally finite} if each pair of vertices is
  related by only finitely many edges.
  A \emph{path} is a finite sequence of
  consecutive edges; the set of paths from~$u$ to~$v$ is defined as
  \[
    \Path(u,v) = \set{
      p\in E^n
      \mid
      n\in \N,
      s(p_1) = u,
      \forall 1 < k \le n.\,t(p_{k-1}) = s(p_{k}),
      t(p_{n}) = v
    }.
  \]
\end{defi}
We often drop the adjectives and simply speak of (multi)graphs. Examples of graphs are visualized later, \eg in \cref{unravelBag}.
All commutative diagrams are also examples of graphs, where each object represents a vertex
and the morphisms represent edges, possibly allowing multiple morphisms between
the same pair of objects.

\begin{exa}\label{exBagCoalg}
  The bag functor $\Bag\colon \Set\to\Set$ sends each set to the set of its finite multisets:
  \[ \Bag X = \set{m\colon X\to \N\mid m(x) = 0\text{ for all but finitely many }x\in X}. \]
  A $\Bag$-coalgebra $c \colon C \rightarrow \Bag C$ is similar to a transition system, with the difference
  that there is a number~$n \in \N$ of transitions from a given state~$x \in C$ to a state~$y \in C$.

  We can identify $\Bag$-coalgebras on~$V$ with locally-finite multigraphs with vertex set~$V$
  by translating between coalgebra structures $c\colon V\to \Bag V$ and edge sets~$E$ as follows:
  \[
    E = \coprod_{\mathclap{(u,v) \in V^2}} \ceil{c(u)(v)}
    \quad
    \setlength{\arraycolsep}{1.4pt}\renewcommand{\arraystretch}{1.3}
    \begin{array}{rcl}
    s\left(\inj_{(u,v)} k\right) &=& u \\
    t\left(\inj_{(u,v)} k\right) &=& v
    \end{array}
    \quad\begin{tikzcd}[row sep=2pt,column sep=8mm]
      {} \arrow[mapsto]{r}
      & {}\\
      {} & \arrow[mapsto]{l}{}\\
    \end{tikzcd}\quad
    c(u)(v) = \big|\set{e\in E\mid s(e) = u, t(e) = v}\big|.
  \]
  From left to right, local finiteness of the graph ensures that
  the definition of~$c(u)(v)$ is a natural number.
\end{exa}

Directed \emph{simple} graphs (with loops) could be modelled as $\Pow$-coalgebras.
Instead, we always think of graphs as coalgebras for the bag functor $\Bag$.
Our motivation is that certain coalgebraic concepts %
yield the expected
graph-theoretic notions in the instantiation for the bag functor, \eg when characterizing trees in the present paper (see \cref{thmBagTree}).
Another such example is colour refinement on graphs
(an early phase in the approximation of graph isomorphism checking
via the Weisfeiler-Leman algorithm~\cite{ShervashidzeSLMB11}),
which is obtained from coalgebraic bisimilarity on $\Bag$-coalgebras~\cite[Ex.~7.18.3]{concurSpecialIssue}.

The bag functor is the prototypical example of an \emph{analytical} functor.
Before stating the definition of analytical functors, we recall some notation.
As usual, given a group~$G$ and a $G$-set~$X$, we denote
the set of orbits of~$X$ under the action of~$G$ by~$X/G$.
This is the quotient of~$X$ by the equivalence relation~$\equiv$ given by
$s \equiv t$ iff $t = g \cdot s$ for some~$g \in G$: \jrnote{}\bknote{}
\[ X/G = \set[\big]{\set{g \cdot t \mid g\in G} \mid t \in X}. \]
If $G$ is a subgroup of the symmetric group~$S_n$ on $n$~elements and $X$ is a set,
then we can define an action of~$G$ on~$X^n$ by $g \cdot t = t \circ g$,
where $\circ$ denotes function composition, and we think of $g$~and~$t$
as functions $n \to n$~and~$n \to X$, respectively.

\begin{defiC}[{\cite{joyal81,joyal86}}]\label{defAnalytical}
  A $\Set$-functor~$F$ is called \emph{analytic} if there is a signature~$\Sigma$ and, for
  each~$\sigma\in \Sigma$, a subgroup~$G_\sigma$ of~$S_{\ar(\sigma)}$ such that
  \[ FX = \coprod_{\sigma\in \Sigma} X^{\ar(\sigma)}/G_\sigma. \]
\end{defiC}
Analytic functors can be understood as polynomial functors for a signature
in which the operation symbols are allowed to forget the order of some of their parameters.
Intuitively, the group~$G_\sigma$ loosens the order on the $\ar(\sigma)$-many
parameters of the operation associated to the symbol~$\sigma$ by
specifying which permutations of a tuple are to be considered equivalent.
The bag functor is the analytic functor with $\Sigma = \N$, $\ar(n) = n$,
and $G_n = S_n$, \ie it has an $n$-ary operation for each $n \in \N$,
which forgets the order of its parameters.\rtnote{}\bknote{}\bknote{}\twnote{}

\begin{defiC}[{\cite[Def.~14.1]{joyofcats}}] \label{D:factSystem}
  Given classes of morphisms $\E$ and $\M$ in $\C$, we say that $\C$ has an
  \emph{$(\E,\M)$-factorization system}\index{EM@$(\E,\M)$}\index{factorization system} provided that:
  \begin{enumerate}
  \item $\E$ and $\M$ are closed under composition and contain all isomorphisms.
    We write
    $\twoheadrightarrow$ for morphisms 
    $e\in \E$, and $\monoto$ for morphisms $m\in \M$.
  \item
    Every morphism $f\colon A\to B$ in $\C$ has a factorization $f =
    m\cdot e$ with $e\in \E$ and $m\in \M$:
    \[
      \begin{tikzcd}[baseline=(A.base),row sep=2mm,column sep=4mm]
        |[alias=A]|
        A
        \arrow[->>,to path={|- (\tikztotarget) \tikztonodes},rounded corners]{dr}[swap,pos=0.3]{e}
        \arrow[]{rr}[alias=f]{f}
        && B
        \\
        & \Im (f)
        \arrow[>->,to path={-| (\tikztotarget) \tikztonodes},rounded corners]{ur}[swap,pos=0.7]{m}
      \end{tikzcd}
    \]
    We write
    $\Im(f)$ for the intermediate object and call it the \emph{image} of $f$.
  \item\label{diagonalization}
    For each commutative square $g\cdot e = m\cdot f$ with $m\in \M$ and $e\in \E$,
    there exists a unique diagonal fill-in $d$ with $m\cdot d=g$ and $d\cdot e =
    f$:
    \[
      \begin{tikzcd}[baseline=(A.base)]
        |[alias=A]|
        A
        \arrow[->>]{r}{e}
        \arrow{d}[swap]{f}
        & B
        \arrow{d}{g}
        \arrow[dashed]{dl}[description]{\exists !d}
        \\
        C
        \arrow[>->]{r}[swap]{m}
        & D
      \end{tikzcd}%
    \]
  \end{enumerate}
  A factorization system is called \emph{proper} if $\E\subseteq \Epi$ and $\M\subseteq \Mono$.
\end{defiC}
Note that in \cref{D:factSystem}, the
classes $\E$ and $\M$ are not assumed to be classes of epis and monos, respectively.

\begin{exas}\hfill
  \begin{enumerate}
  \item
  $\Set$ has an $(\Epi,\Mono)$-factorization system.
  \item Every category $\C$ has an $(\Iso,\Mor)$-factorization system,
  which is not proper in general.
  \end{enumerate}
\end{exas}

\begin{defi}
For a class $\M$ of morphisms in $\C$ and a functor $F \colon \C \to \C$,
let $\liftfactsys{\M}{F}$ be the class of those (pointed) $F$-coalgebra morphisms $h \colon (C,c) \to (D,d)$
with $h \in \M$.
\end{defi}
\begin{propC}[{\cite{Wissmann22}}]\label{coalgFact}
If $F \colon \C \to \D$ preserves $\M$ (that is, $Fm \in \M$ for every $m \in \M$),
then every\footnote{Note that this includes non-proper
factorization systems; earlier proofs, \eg by Kurz~\cite[1.3.5-1.3.7]{kurzPhd}, impose
restrictions on $\E$ and $\M$.} $(\E,\M)$-factorization system of $\C$ lifts to an
$(\liftfactsys{\E}{F}, \liftfactsys{\M}{F})$-factorization system on $\Coalg(F)$ and on $\Coalg_I(F)$.
\end{propC}

\begin{rem}\label{remCancel}
  Factorization systems have cancellation and stability
  properties similar to those satisfied by monomorphisms and epimorphisms. For instance,
  if a composite $g\cdot f$ is in $\M$ and $g$ is in $\M$, then $f$ is in $\M$,
  too~\cite[Prop.~14.9(1)]{joyofcats} (in the case of properness, the
  assumption $g\in \M$ can be dropped).
  Furthermore, $\M$-morphisms are stable under pullback~\cite[Prop.~14.15(2)]{joyofcats}.
\end{rem}
\begin{defi}
  We refer to the pullback of an $\M$-morphism along an $\M$-morphism as a
  (binary) \emph{$\M$-intersection}.
\end{defi}
\begin{exa}\label{exIntersection}
  If $\M$ is a class of monomorphisms in $\Set$, then an $\M$-intersection is an intersection of subobjects in the usual sense.
  If $\M=\Mor$, \ie $\M$ is the class of all morphisms in $\C$, then an $\M$-intersection is a (general) pullback in $\C$.
\end{exa}
\begin{exa}\label{exCoalgIntersection}
  Pointed $F$-coalgebras also have $\M$-intersections under mild conditions:
  if the base category $\C$ has $\M$-intersections and if $F$ preserves them
  weakly (\ie it maps pullback squares of $\M$-morphisms to weak pullback squares), then
  $\Coalg[I](F)$ has pullbacks of $\liftfactsys{\M}{F}$-morphisms.
  \begin{enumerate}
  \item For $\C=\Set$ and $\M\subseteq \Mono$, every functor $\Set\to\Set$
  preserves binary intersections up to redefinition of
  $F\emptyset$~\cite{trnkova71,AT90,Barr93} (see~\cite{wmkd20reachability} for
  a survey). This redefinition leaves $\Coalg[I](F)$ untouched, because there is
  at most one pointed coalgebra on $\emptyset$ (and none if $I\neq \emptyset$).

  \item For $\C=\Set$ and $\M=\Mor$, this preservation condition boils down to the weak preservation of pullbacks. This
  criterion is studied at length in the coalgebraic literature~\cite{Gumm2005},
  as this is the standard sufficient condition for coalgebraic bisimilarity and
  coalgebraic behavioural equivalence to coincide.
  \end{enumerate}
\end{exa}

\noindent We shall also need the notions of \emph{wide} pullbacks and their preservation.\bknote{}
\begin{defi}\hfill
  \begin{enumerate}
  \item A \emph{wide pullback} of a family $(f_i \colon A_i \to X)_{i \in I}$
  is a limiting cone for the diagram consisting of the object~$X$ and all the morphisms~$f_i$.
  (The object~$X$ is still part of the diagram even if $I$ is empty.)
  Spelling this out, a wide pullback of~$(f_i)_{i \in I}$ consists of an object~$P$,
  a morphism $g \colon P \to X$, and a family $(p_i \colon P \to A_i)_{i \in I}$
  such that $f_i \cdot p_i = g$ for all~$i \in I$, and which is universal in the sense that
  for any object~$P'$, morphism $g' \colon P' \to X$, and family $(p'_i \colon P' \to A_i)_{i \in I}$
  with $f_i \cdot p'_i = g'$ for all~$i \in I$, there exists a unique morphism $h \colon P' \to P$
  such that $g \cdot h = g'$ and $p_i \cdot h = p'_i$ for all~$i \in I$.
  \item A \emph{weak wide pullback} is defined similarly, but without the uniqueness requirement on~$h$.
  \item We refer to a wide pullback of a family of $\M$-morphisms as a \emph{wide $\M$-intersection}. \emph{Weak wide $\M$-intersections} are defined similarly. 
  \item A functor~$F$ \emph{weakly preserves} wide $\M$-intersections if it maps wide $\M$-intersections to weak wide pullbacks, \ie if for every family $(m_i \colon M_i \monoto X)_{i \in I}$
  of $\M$-morphisms and any wide pullback $(C, g \colon C \to X, c_i \colon C \to M_i)_{i \in I}$ of~$(m_i)_{i \in I}$,
  the cone
  \[ (FC, Fg \colon FC \to FX, Fc_i \colon FC \to FM_i)_{i \in I} \]
  is a weak wide pullback of
  the family $(Fm_i \colon FM_i \to FX)_{i \in I}$.
  Similarly, $F$ \emph{preserves weak wide $\M$-intersections} if it maps weak wide $\M$-intersections to weak wide pullbacks.
  \end{enumerate}
\end{defi}
\begin{rem}\label{remStableWidePullback}
  Besides binary pullbacks, $\M$-morphisms of an $(\E,\M)$-factorization system
  are also closed under wide pullbacks~\cite[Prop.~14.15]{joyofcats}. This means
  that if $(C,g,c_i)_{i \in I}$ is a wide pullback of~$(m_i)_{i \in I}$ and
  $m_i \in \M$ for all~$i \in I$, then $g \in \M$.
\end{rem}

\begin{lem}\label{lemIntsecMonoPres}
  Suppose $\M \subseteq \Mono$ and $F \colon \C \to \D$ weakly preserves $\M$-intersections.
  Then $F$ sends $\M$ morphisms to monos.
\end{lem}
\begin{proof}
  It is well-known that a morphism $f \colon A \to B$ is monic iff
  \begin{equation}\label{eq:monoSquare}
    \begin{tikzcd}
      A
      \arrow{r}{\id}
      \arrow{d}[swap]{\id}
      & B
      \arrow{d}{f}
      \\
      A
      \arrow{r}[swap]{f}
      & B
    \end{tikzcd}
  \end{equation}
  is a pullback square. Since $\id$ is monic, this is the case already if it is a \emph{weak} pullback square.
  Now, if $m \in \M \subseteq \Mono$, then \labelcref{eq:monoSquare} with $f := m$ is a pullback square.
  By assumption, since $m \in \M$, $F$ weakly preserves this square, so \labelcref{eq:monoSquare} with $f := Fm$ is a weak pullback square.
  This then implies that $Fm$ is a mono.
\end{proof}

\section{Precise Morphisms}\label{sec:precise}
Towards a description of trees by an iterative construction on coalgebras, we start
by discussing the existing notion of \emph{precise morphisms}~\cite[Def.~3.1]{WDKH2019}
that intuitively describes the relation between two \emph{levels} of a
tree: every node further down is the child of precisely one node in the level above.
\begin{defiC}[{\cite[Def.~3.1]{WDKH2019}}]\label{defPrecise}
  For a functor $F\colon \C\to \D$ and a class~$\M$ of morphisms in~$\C$, a
  morphism $p\colon P\to FR$ in $\D$ is called \emph{$F$-precise} or simply \emph{precise} (\wrt $\M$)
  if for all~$g \colon P \to FC$ and for all $m \colon R \to D$ and $n \colon C \to D$ in~$\M$,
  the following implication holds:
  \[
    \begin{tikzcd}[sep=6mm,column sep=12mm]
      P
      \arrow{d}[swap]{g}
      \arrow{r}{p}
      & FR
      \arrow{d}{Fm}
      \\
      FC
      \arrow{r}[swap]{Fn}
      & FD
    \end{tikzcd}
    \quad\overset{\exists d\,}{\Longrightarrow}\quad
    \begin{tikzcd}[sep=6mm]
      P
      \arrow{d}[swap]{g}
      \arrow{r}{p}
      & FR
      \arrow{dl}{Fd}
      \\
      FC
    \end{tikzcd}
    \&
    \begin{tikzcd}[sep=6mm]
      & R
      \arrow[>->]{d}{m}
      \arrow{dl}[swap]{d}
      \\
      C
      \arrow[>->]{r}[swap]{n}
      & D
    \end{tikzcd}
  \]
\end{defiC}

If $\M$ is part of an $(\E,\M)$-factorization system, then by $m,n\in \M$, the diagonal $d$ is
necessarily in $\M$ as well (\cref{remCancel}). The present definition is a mild generalization of
the original definition~\cite{WDKH2019}, which can be reobtained by setting
$\C=\D$ and setting $\M$ to be the class of \emph{all} morphisms.

\begin{exaC}[{\cite{WDKH2019}}]
  \label{exPreciseMap}
  For intuition, we consider the simple case where $\M$ is the class of all
  morphisms and $\C=\D=\Set$.
  For a polynomial $\Set$-functor $F$, a map $p\colon P\rightarrow
  FR$ is $F$-precise iff every element of $R$ is mentioned precisely once in the
  definition of the map $p$. In particular, for the functor $\FBinTree$ from \itemref{ex:coalgebras}{ex:bintree},
  a map $p\colon P\to \FBinTree R$ is precise iff
  \[
    \forall y\in R,\,\exists! (x,i)\in P\times \set{1,2}\colon p(x) = (y_1,y_2)\text{ and } y_i = y.
  \]
  The map $f\colon X\to \FBinTree Y$ in
  \cref{figExPrecise} is not $\FBinTree$-precise, because $y_2$ is used three times
  (once in $f(x_2)$ and twice in $f(x_3)$), and $y_3$ and $y_4$ do not occur in
  $f$ at all. The map $p\colon X\to \FBinTree Y'$ is $\FBinTree$-precise because every element
  of $Y'$ is used precisely once in $p$. The equality $\FBinTree h\cdot p = f$ witnesses
  that $f$ is indeed not precise. To see this, consider the commutative
  square $\FBinTree h\cdot p = \FBinTree\id_{Y}\cdot f$. There is no
  diagonal $d\colon Y\to Y'$ with $h\cdot d = \id_{Y}$ (nor one with $Fd \cdot f = p$).
  Thus, $f$ does not satisfy the definition of precise maps (\cref{defPrecise})
  for $g := p, n := h$, and $m := \id_Y$.
\end{exaC}

\begin{figure}[h]\centering%
  \begin{tikzpicture}[xscale = 1,yscale=0.6]
    \foreach \prefix/\varname/\labeltext/\count/\xshift in
    {x/x/X/4/0,y/y/Y/4/2,xcopy/x/X/4/6,
      yprime/{\ensuremath{y'}}/{\ensuremath{Y'}}/4/8,ycopy/y/Y/4/10} {
      \foreach \n in {1,...,\count} {
        \node[named vertex] (\prefix\n) at (\xshift cm,-\n) {$\varname_\n$};
      }
      \node[set,fit=(\prefix1) (\prefix\count),label={[name=\prefix]$\labeltext$}] (\prefix domain) {};
    }
    \path[myedge] (x2) to (y1) ;
    \path[myedge] (x2) to[bend left=20] (y2.north west) ;
    \path[myedge] (x3) to[bend left] (y2) ;
    \path[myedge] (x3) to[bend right] (y2) ;
    \foreach \node in {x1,x4,xcopy1,xcopy4} {
      \path[myedge] (\node) to node[pos=1]{$\bot$} +(8mm,0) ;
    }
    \path[myedge] (xcopy2) to (yprime1) ;
    \path[myedge] (xcopy2) to (yprime2) ;
    \path[myedge] (xcopy3) to (yprime3) ;
    \path[myedge] (xcopy3) to (yprime4) ;
    \path[myedge] (yprime1) to (ycopy1);
    \path[myedge] (yprime2) to[bend left=15] (ycopy2);
    \path[myedge] (yprime3) to[bend right=0] (ycopy2);
    \path[myedge] (yprime4) to[bend right=18] (ycopy2);
    \path[every label]
    (x) edge[draw=none] node {$f$} (y)
    (xcopy) edge[draw=none] node {$p$} (yprime)
    (yprime) edge[draw=none] node {$h$} (ycopy)
    ;
    \node[anchor=base east] at ([xshift=-4mm]x4.base) {\(
    \begin{array}[b]{r@{\,}c@{\,}l}
      f\colon X & \to & \FBinTree Y \\
      x_1 & \mapsto & \bot \\
      x_2 & \mapsto & (y_1,y_2) \\
      x_3 & \mapsto & (y_2,y_2) \\
      x_4 & \mapsto & \bot \\
    \end{array}
    \)};
    \node[anchor=base east] at ([xshift=-4mm]xcopy4.base) {\(
    \begin{array}[b]{r@{\,}c@{\,}l}
      p\colon X & \to & \FBinTree Y' \\
      x_1 & \mapsto & \bot \\
      x_2 & \mapsto & (y_1',y_2') \\
      x_3 & \mapsto & (y_3',y_4') \\
      x_4 & \mapsto & \bot \\
    \end{array}
    \)};
  \end{tikzpicture}
  \caption{A non-precise map $f$ that factors through the $\FBinTree$-precise $p$
  for $\FBinTree X = X\times X+\set{\bot}$~\cite{WDKH2019}.}
  \label{figExPrecise}
\end{figure}

In contrast to \cref{exPreciseMap}, there are essentially no precise maps for the powerset functor.
The intuitive reason is that
for $x\in X$, the sets
\[
  \set{x}
  =
  \set{x,x}
  = \set{x,x,x}
  = \cdots
\]
all describe the same element in $\Pow X$, so
one can consider the element $x$ to have
unbounded multiplicity in $\set{x}$.
The following proposition formalizes this intuition:
\begin{prop}\label{powPrecise}
A map $f\colon X\to \Pow Y$ is $\Pow$-precise \wrt $\Mor$ iff $Y = \emptyset$.
\end{prop}
\begin{proof}
Even though the idea is sketched in the appendix of~\cite[Ex.~3.7]{WDKH2019},
we provide a more detailed argument here for the convenience of the reader.

For the `if' direction, let $Y = \emptyset$, and let $g \colon X \to PZ$, $m \colon Y \to A$,
and $n \colon Z \to A$ such that $\Pow m \cdot f = \Pow n \cdot g$. Since $f(x) = \emptyset$
for all $x \in X$, we necessarily have $g(x) = \emptyset$ as well for all $x \in X$. Thus, the unique map
$d \colon Y \to Z$ makes the required two triangles commute.

For `only if', suppose $f \colon X \to \Pow Y$ is precise.
Let $g \colon X \to \Pow(Y + Y)$ be defined as
$g(x) = (\Pow \inl)(f(x)) \cup (\Pow \inr)(f(x))$.
Choosing $m = \id_Y$ and $n = \nabla_Y = [\id_Y, \id_Y] \colon Y + Y \to Y$, we have
$\Pow \id_Y \cdot f = \Pow \nabla_Y \cdot g$. By preciseness of $f$, there is a $d \colon Y \to Y + Y$
such that $\Pow d \cdot f = g$ and $\nabla_Y \cdot d = \id_Y$. The second equation
implies that $d(y) \in \{\inl(y), \inr(y)\}$ for all $y \in Y$.

All in all, we can show that $f(x) = \emptyset$ for all $x \in X$:
For let $x \in X$ and suppose $y \in f(x)$.
Then either $\inl(y)\in (\Pow d)(f(x))$ or $\inr(y)\in
(\Pow d)(f(x))$, but not \emph{both} (by $\Pow d$). But by the definition of $g$, we have both
$\inl(y)\in g(x) = (\Pow d)(f(x))$ and $\inr(y)\in g(x) = (\Pow d)(f(x))$, a contradiction.

Thus, we have the commutative square
(with the unique map $X\to \Pow\emptyset$):
\[
  \begin{tikzcd}
    X
    \arrow{r}{f}
    \arrow{d}[swap]{!}
    & \Pow Y
    \arrow{d}{\Pow \id_Y}
    \\
    \Pow \emptyset
    \arrow{r}[swap]{\Pow i}
    & \Pow Y
  \end{tikzcd}
\]
Since $f$ is $\Pow$-precise, we obtain a diagonal $d\colon Y\to \emptyset$, showing that $Y=\emptyset$.
\end{proof}

We remedy the fact that $\Pow$ is not well-behaved with respect to precise maps
by using the bag functor~$\Bag$ instead of~$\Pow$.
As for polynomial functors (see \cref{exPreciseMap}),
a map~$f \colon X \to \Bag Y$ is $\Bag$-precise iff every element of~$Y$
is mentioned precisely once in the definition of~$f$:
\begin{propC}[{\cite[Lem.~A.6]{WDKH2019}}]
  A morphism $f \colon X \to \Bag Y$ is $\Bag$-precise \wrt $\Mor$ iff
  for every~$y \in Y$, we have
  \[ \sum_{x \in X} f(x)(y) = 1. \]
\end{propC}
\begin{exa}
  For the bag functor, the induced diagonal in \cref{defPrecise}
  is not necessarily unique. For instance, $f\colon 1\to \Bag 2$, $f(*)(x)=1$ is $\Bag$-precise.
  However, both the identity $\id_2$ and the swap function $s\colon 2\to 2$ are
  diagonals in the following commutative square:
  \[
  \begin{tikzcd}
    1
    \arrow{r}{f}
    \arrow{d}[swap]{f}
    & \Bag 2
    \arrow{dl}[above left]{\Bag \id_2}
    \arrow{d}{\Bag !}
    \\
    \Bag 2
    \arrow{r}{\Bag !}
    & \Bag 1
  \end{tikzcd}
  \quad
  \begin{tikzcd}
    1
    \arrow{r}{f}
    \arrow{d}[swap]{f}
    & \Bag 2
    \arrow{dl}[above left]{\Bag s}
    \arrow{d}{\Bag !}
    \\
    \Bag 2
    \arrow{r}{\Bag !}
    & \Bag 1
  \end{tikzcd}
  \]
  Coincidentally, the two diagonals $\id_2$ and $s$ are precisely the two elements of the symmetric group $S_2$,
  which is used to describe the bag functor as an analytical
  functor (\cref{defAnalytical}), that is, as a quotient of $X\mapsto \coprod_{n\in \N} X^n$.
\end{exa}

\begin{rem}
	The definition of precise morphisms (\cref{defPrecise}) looks similar to that of
$(\E,\M)$-functors~\cite[Chp.~17]{joyofcats}, but differs in the following
aspects:
\begin{enumerate}
\item For precise morphisms, both $m$ and $n$ are required to be in
$\M$, which is crucial for the unification with least bounds, as discussed next.
Instead, in the diagram for $(\E,\M)$-functors,
only the \textqt{opposite side} $n$ is required to be in $\M$.

\item In contrast to $(\E,\M)$-functors, the diagonal in \cref{defPrecise} is
not necessarily unique.

\item The factorization of $(\E,\M)$-functors is defined in terms of
  \emph{sources}, that is, families of morphisms with a common domain.
\end{enumerate}
\end{rem}

\noindent Under mild conditions, the morphism $m$ in \cref{defPrecise}
can be assumed to be the identity. Concretely:
\begin{lem}\label{preciseWithIntersections}
  If $\C$ has an $(\E,\M)$-factorization system, $\C$ has $\M$-intersections, and $F$ weakly preserves $\M$-intersections, then
  a morphism $p\colon P\to FR$ in~$\D$ is $F$-precise \wrt $\M$ iff for
  all $g\colon P\to FC$ and for all $n\colon C\to R$ in~$\M$, the following
  implication holds:
  \[
    \begin{tikzcd}[sep=6mm,column sep=12mm]
      P
      \arrow{d}[swap]{g}
      \arrow{r}{p}
      & FR
      \\
      FC
      \arrow{ur}[swap,pos=0.4]{Fn}
    \end{tikzcd}
    \quad\overset{\exists d\,}{\Longrightarrow}\quad
    \begin{tikzcd}[sep=6mm]
      P
      \arrow{d}[swap]{g}
      \arrow{r}{p}
      & FR
      \arrow{dl}{Fd}
      \\
      FC
    \end{tikzcd}
    \&
    \begin{tikzcd}[sep=6mm]
      & R
      \arrow[>->]{d}{\id_R}
      \arrow{dl}[swap]{d}
      \\
      C
      \arrow[>->]{r}[swap]{n}
      & R
    \end{tikzcd}
  \]
\end{lem}
\begin{proof}
  The proof can be adapted straightforwardly from the case where
  $\M$ is the class of all morphisms~\cite[Rem.~3.2]{WDKH2019}:
  \begin{itemize}
  \item For \textqt{only if}, let $p\colon P\to FR$ be precise and consider
  $g\colon P\to FC$ and $n\in \M$ with $p = Fn\cdot g$. Since $\M$ is part of a factorization system,
  we have $\id_R\in \M$. Hence, since $p$ is precise,
  we can put $D := R$ and $m := \id_R$ in \cref{defPrecise} and obtain a diagonal $d\colon R\to C$ with
  $Fd\cdot p = g$ and $n\cdot d = \id_R$, as desired.

  \item For \textqt{if}, consider a commutative square
    $Fm\cdot p = Fn\cdot g$ (as in \cref{defPrecise}) with $m,n\in \M$, and form the pullback $B$ of $m$ and $n$:
  \[
    \begin{tikzcd}[sep=6mm]
      B
      \arrow{d}[swap]{m'}
      \arrow{r}{n'}
      \pullbackangle{-45}
      & R
      \arrow{d}{m}
      \\
      C
      \arrow{r}[swap]{n}
      & D
    \end{tikzcd}
    \quad\Longrightarrow\quad
    \begin{tikzcd}[sep=6mm]
      FB
      \arrow{d}[swap]{Fm'}
      \arrow{r}{Fn'}
      & FR
      \arrow{d}{Fm}
      \\
      FC
      \arrow{r}[swap]{Fn}
      & FD
    \end{tikzcd}
  \]
  Since $F$ preserves pullbacks weakly, the right hand side of the above diagram is a weak pullback.
  Hence, the commuting square $Fm \cdot p = Fn \cdot g$ induces some (not necessarily unique) morphism $g'\colon P\to FB$ making the two triangles in the diagram below commute:
  \begin{equation}
    \begin{tikzcd}[sep=6mm,column sep=12mm]
      P
      \arrow{dd}[swap]{g}
      \arrow{rr}{p}
      \arrow[dashed]{dr}[swap]{g'}
      && FR
      \arrow{dd}{Fm}
      \\
      & FB
      \arrow{ru}[swap,pos=0.4]{Fn'}
      \arrow{dl}[swap,pos=0.3]{Fm'}
      \\
      FC
      \arrow{rr}[swap]{Fn}
      && FD
    \end{tikzcd}
    \label{eqSimpleDiagStart}
  \end{equation}
  Furthermore, since $\M$-morphism are stable under pullback (\cref{remCancel}), we have $n' \in \M$.
  Thus, we can apply our assumption to $p = Fn'\cdot g'$
  to obtain $d\colon R\to B$ making the following diagrams commute:
  \[
    \begin{tikzcd}[sep=6mm]
      P
      \arrow{d}[swap]{g'}
      \arrow{r}{p}
      & FR
      \arrow[dashed]{dl}{Fd}
      \\
      FB
    \end{tikzcd}
    \&
    \begin{tikzcd}[sep=6mm]
      & R
      \arrow[>->]{d}{\id_R}
      \arrow[dashed]{dl}[swap]{d}
      \\
      B
      \arrow[>->]{r}[swap]{n'}
      & R
    \end{tikzcd}
  \]
  The morphism $m'\cdot d\colon R\to C$ is then the desired diagonal fill-in
  for the original commutative square~\labelcref{eqSimpleDiagStart},
  as witnessed by the following commuting diagrams:
  \[
    \begin{tikzcd}[sep/.append code=\pgfkeyssetvalue{/tikz/sep}{#1},
      sep=6mm,baseline=(\tikzcdmatrixname-3-1.base)]
      P
      \arrow{dr}[swap,pos=0.6]{g'}
      \arrow{rr}{p}
      \arrow{dd}[swap]{g}
      && FR
      \arrow{dl}{Fd}
      \\
      & FB
      \arrow{dl}{Fm'}
      &&[-\pgfkeysvalueof{/tikz/sep}]\&
      \\
      FC
    \end{tikzcd}
    \begin{tikzcd}[sep=6mm,baseline=(\tikzcdmatrixname-3-1.base)]
      & & R
      \arrow[>->]{d}[pos=0.6]{\id_R}
      \arrow{dl}[swap]{d}
      \\
      & B
      \arrow[>->]{r}[swap]{n'}
      \arrow{dl}[swap]{m'}
      & R
      \arrow[>->]{d}{m}
      \\
      C
      \arrow[>->]{rr}[swap]{n}
      && D
    \end{tikzcd}
    \qedhere
  \]
  \end{itemize}
\end{proof}
\begin{rem}
If, in addition to the conditions of \cref{preciseWithIntersections}, we also have
$\M\subseteq \Mono$, then $p$ is precise iff for any $g$ and $n \in \M$, $p =
Fn \cdot g$ implies that $n$ is an isomorphism:
  \begin{itemize}
  \item For \textqt{only if}, let $p\colon P\to FR$ be precise and consider
  $g\colon P\to FC$ and $n\in \M$ with $p = Fn\cdot g$. Since $p$ is precise,
  we obtain a diagonal $d\colon R\to C$ with $n\cdot d = \id_R$. Thus, $n$ is
  monic and a split epi, and hence an isomorphism.

  \item For \textqt{if}, in order to check that $p\colon P\to FR$ is precise,
  consider $g\colon P\to FC$ and $n\in \M$ with $p = Fn\cdot g$. By assumption,
  $n$ is an isomorphism, and so its inverse $d\colon C\to R$ satisfies $n\cdot
  d = \id_R$ and $Fd\cdot p = Fd\cdot Fn \cdot g = g$, as desired.
  \end{itemize}
  This characterization looks akin to the definition of extremal
  epimorphisms~\cite[Def.~7.74]{joyofcats}, only with a functor $F$ inserted.
  In fact, picking $F = \Id$ and $\M = \Mono$, we obtain that in a category with intersections,
  $\Id$-precise morphisms and extremal epimorphisms coincide.
\end{rem}

For right adjoint functors, we can characterize precise morphisms in terms of their adjoint transposes:
\begin{prop}\label{rightAdjointPrecise}
  Suppose that $\C$ has an $(\E, \M)$-factorization system and that $F\colon \C\to \D$ has a left adjoint~$L$.
  Then a morphism $p \colon P \to FR$ is $F$-precise \wrt $\M$ iff
  its adjoint transpose $p^\# \colon LP\to R$ is in~$\E$.
  Moreover, if $p$ is precise, then its diagonal fill-ins are unique.
\end{prop}
\begin{proof}
  ($\Rightarrow$)
  Given that $p\colon P\to FR$ is precise, take the
  $(\E, \M)$-factorization of its adjoint transpose $p^\#\colon LP\to R$:
  \[
  \begin{tikzcd}
    LP
    \arrow[->>]{r}{e}
    \arrow[shiftarr={yshift=6mm}]{rr}{p^\#}
    & Q
    \arrow[>->]{r}{m}
    & R
  \end{tikzcd}
  \]
  The adjoint transpose $\bar e$ of $e$ satisfies $p=Fm\cdot \bar e$.
  Hence, using \cref{preciseWithIntersections}, we obtain
  a diagonal making the following triangles commute:
  \[
    \begin{tikzcd}[sep=6mm]
      P
      \arrow{d}[swap]{\bar e}
      \arrow{r}{p}
      & FR
      \arrow[dashed]{dl}{Fd}
      \\
      FQ
    \end{tikzcd}
    \&
    \begin{tikzcd}[sep=6mm]
      & R
      \arrow[>->]{d}[pos=0.6]{\id_R}
      \arrow[dashed]{dl}[swap]{d}
      \\
      Q
      \arrow[>->]{r}[swap]{m}
      & R
    \end{tikzcd}
  \]
  Transposing the left hand triangle, we get $d \cdot p^\# = e$.
  The dual of~\cite[Lem.~14.5]{joyofcats} states that in any commutative square of the form
  \[
    \begin{tikzcd}
      \bullet
      \arrow[<-<]{r}{}
      \arrow[<-]{d}[swap,pos=0.6]{\id}
      &
      \bullet
      \arrow[<-]{dl}
      \arrow[<<-]{d}
      \\
      \bullet
      \arrow[<-]{r}[swap]{q}
      & \bullet
    \end{tikzcd}
  \]
  we necessarily have $q\in \E$. And indeed, we have the following commutative
  diagram for $q := p^\#$:
  \[
    \begin{tikzcd}
      R
      \arrow[<-<]{r}{m}
      \arrow[<-]{d}[swap,pos=0.6]{\id_R}
      &
      Q
      \arrow[<-]{dl}[description]{d}
      \arrow[<<-]{d}[pos=0.6]{e}
      \\
      R
      \arrow[<-]{r}[swap]{p^\#}
      & LP
    \end{tikzcd}
  \]
  It follows that $p^\#\in \E$, as desired.

  ($\Leftarrow$)
  Suppose that $p^\# \in \E$ and 
  consider a commutative square
  \begin{equation}\label{eq:radj-square}
    \begin{tikzcd}[sep=6mm,column sep=12mm]
      P
      \arrow{d}[swap]{g}
      \arrow{r}{p}
      & FR
      \arrow{d}{Fm}
      \\
      FC
      \arrow{r}[swap]{Fn}
      & FD
    \end{tikzcd}
  \end{equation}
  with $m,n \in \M$. The adjunction yields the outer commutative square in the diagram below:
  \begin{equation}\label{eq:radj-transp-square}
    \begin{tikzcd}[sep=6mm,column sep=12mm]
      LP
      \arrow{d}[swap]{g^\#}
      \arrow[->>]{r}{p^\#}
      & R
      \arrow[>->]{d}{m}
      \arrow[dashed]{dl}[description]{d}
      \\
      C
      \arrow[>->]{r}[swap]{n}
      & D
    \end{tikzcd}
  \end{equation}
  Applying the diagonal fill-in property to this square, we get a morphism $d$ satisfying $n\cdot d = m$ and $d\cdot p^\# = g^\#$. Transposing the second equality along the adjunction again yields the remaining desired equality $Fd\cdot p = g$.

  For the final claim about uniqueness, suppose we have two diagonals $d$ and $d'$ for the square~\labelcref{eq:radj-square}.
  Transposing along the adjunction yields that both $d$ and $d'$ are fill-ins for the square~\labelcref{eq:radj-transp-square}.
  Hence, the uniqueness of diagonals in factorization systems gives us $d = d'$.
\end{proof}

\begin{exa}\label{idPrecise}
  For every set~$Y$, we have an adjunction $(-)\times Y \dashv (-)^Y$.
  By \cref{rightAdjointPrecise}, a map
  $p\colon P\to R^Y$ is $(-)^Y$-precise \wrt $\Mor$ iff its
  uncurried version $p^\#\colon P\times Y\to R$ is a bijection.
  Moreover, since the identity functor is right adjoint to itself,
  we obtain that in a category~$\C$ with an $(\E, \M)$-factorization
  system, a morphism $p \colon P \to R$ is $\Id_\C$-precise \wrt $\M$ iff $p \in \E$.
\end{exa}

\subsection{Precise Factorizations}
As illustrated in \cref{figExPrecise}, certain functors allow factorization of
morphisms of the shape $X\to FY$ through precise maps.

\begin{defiC}[{\cite[Def.~3.4]{WDKH2019}}] \label{defPreciseFactor}
  We say
  that $F \colon \C \to \D$ \emph{admits precise factorizations (\wrt $\M$)} if for every
  $f\colon P\rightarrow FY$, there exist an $F$-precise morphism
  $p\colon P\rightarrow FR$ and
  a morphism $h\colon R\monoto Y$ in~$\M$ with $Fh\cdot p = f$:
  \[
    \begin{tikzcd}
      P
      \arrow{dr}[swap]{f}
      \arrow[dashed]{r}{p}
      & FR
      \arrow[dashed]{d}{Fh}
      \\
      &  FY
    \end{tikzcd}
  \]
  The triple~$(R,p,h)$ is called the \emph{precise factorization} of~$f$.
\end{defiC}

The original definition~\cite{WDKH2019}
is for endofunctors and parametric in a class $\S$ of objects (requiring $P, R \in \S$), which we omit in
the present paper for the sake of simplicity.
In any case, even though the diagonal in the definition of precise morphism is
not required to be unique, we can still deduce uniqueness of precise factorizations:

\begin{lem}\label{preciseUnique}
  If $\M$ is part of an $(\E,\M)$-factorization system, then precise
  factorizations are unique up to isomorphism.
  Concretely, if $p_1$ and $p_2$ are precise and $g,h\in \M$ satisfy $Fg\cdot p_1 = Fh\cdot p_2$,
  then the induced morphism $d$ is an isomorphism:
  \[
    \begin{tikzcd}
      X
      \arrow{r}{p_1}
      \arrow{d}[swap]{p_2}
      & FY_1
      \arrow[dashed]{dl}[description]{\cong}[above left,xshift=-1mm,yshift=1mm]{Fd}
      \arrow{d}{Fg}
      \\
      FY_2
      \arrow{r}[swap]{Fh}
      & FZ
    \end{tikzcd}%
  \]
\end{lem}
\begin{proof}
  The original uniqueness proof~\cite[Rem.~3.5]{WDKH2019} extends to our setting for a morphism class $\M$.
  Actually, we only need that $\M$ contains every identity morphism and
  that $\M$ satisfies the cancellation property of \cref{remCancel}.
  By this cancellation property, we obtain $d\in \M$ from $h\cdot d = g$.
  Since $\id_{Y_2} \in \M$, the preciseness of $p_2$ yields a morphism $s$:
  \[
    \begin{tikzcd}[column sep=12mm]
      X
      \arrow{r}{p_2}
      \arrow{d}[swap]{p_1}
      & FY_1
      \arrow{d}[swap]{F\id_{Y_2}}{~(\id_{Y_2}\in \M)}
      \\
      FY_2
      \arrow{r}{Fd}[swap]{(d\in \M)}
      & FY_2
    \end{tikzcd}
    \quad\overset{\exists s}{\Longrightarrow}\quad
    \begin{tikzcd}
      X
      \arrow{r}{p_2}
      \arrow{d}[swap]{p_1}
      & FY_1
      \arrow{d}{F\id_{Y_2}}
      \arrow[dashed]{dl}[description]{Fs}
      \\
      FY_2
      \arrow{r}[swap]{Fd}
      & FY_2
    \end{tikzcd}
    \text{ with }d\cdot s = \id_{Y_2}
  \]
  Again by cancellation, $s\in \M$.
  Since $\id_{Y_1}\in \M$, take the following commutative square and apply that $p_1$ is precise:
  \[
    \begin{tikzcd}[column sep=12mm]
      X
      \arrow{r}{p_1}
      \arrow{d}[swap]{p_2}
      & FY_1
      \arrow{d}[swap]{F\id_{Y_1}}{~(\id_{Y_1}\in \M)}
      \\
      FY_2
      \arrow{r}{Fs}[swap]{(s\in \M)}
      & FY_1
    \end{tikzcd}
    \quad\overset{\exists t}{\Longrightarrow}\quad
    \begin{tikzcd}[column sep=12mm]
      X
      \arrow{r}{p_1}
      \arrow{d}[swap]{p_2}
      & FY_1
      \arrow{d}{F\id_{Y_1}}
      \arrow[dashed]{dl}[description]{Ft}
      \\
      FY_2
      \arrow{r}[swap]{Fs}
      & FY_1
    \end{tikzcd}
    \text{ with }
    s\cdot t = \id_{Y_1}
  \]
  In total, $s$ has both a left-inverse $d$ and a right inverse $t$. Thus, $d =
  t$~\cite[Prop.~3.10]{joyofcats}, showing that $d$ is an isomorphism.\bknote{}%
\end{proof}

Many functors admit precise factorizations:

\begin{prop}\label{rightAdjointPreciseFact}
  Every right-adjoint functor $F\colon \C\to \D$ from a category $\C$ with an $(\E,\M)$-factorization system admits
  precise factorizations \wrt $\M$.
\end{prop}
\begin{proof}
  Denote the left adjoint of $F$ by $L$.
  For $f\colon P\to FY$, consider the $(\E,\M)$-factorization of its adjoint transpose $f^\#\colon LP\to Y$:
  \[
    \begin{tikzcd}
      LP
      \arrow[->>]{r}{e}
      \arrow[shiftarr={yshift=6mm}]{rr}{p^\#}
      &
      R
      \arrow[>->]{r}{m}
      & Y
    \end{tikzcd}
  \]
  Transposing this diagram, we get $Fm\cdot \bar e = p$, where $\bar e \colon P \to FR$ is the adjoint transpose of $e$.
  Moreover, the morphism $m$ is in $\M$, and $\bar e$ is $F$-precise by \cref{rightAdjointPrecise}.
  Hence, $(R, \bar e, m)$ is the precise factorization of $p$.
\end{proof}

\begin{cor}\label{idPreciseFact}
Suppose $\C$ has an $(\E, \M)$-factorization system.
Then $\Id_\C$ admits precise factorizations \wrt $\M$.
\end{cor}
\begin{proof}
  Consequence of \cref{rightAdjointPreciseFact} since $\Id_C$ is right adjoint to itself.
\end{proof}

\begin{propC}[{\cite[Prop.~3.6]{WDKH2019}}]\label{morEndoPrecise}
  The following functors admit precise factorizations \wrt $\Mor$:
  \begin{enumerate}
  \setlength{\itemsep}{1pt plus 1pt}
  \setlength{\parskip}{0pt}
  \item\label{constantPrecise} Constant functors $\C \to \D$, if $\C$ has an initial object.
  \item $F\cdot F'$, if $F \colon \C \to \D$ and $F' \colon \mathcal{B} \to \C$ do so.
  \item $\prod\limits_{i\in I} F_i$, if all $(F_i \colon \C \to \D)_{i\in I}$ do so
    and $\C$ has $I$-indexed coproducts.
  \item $\coprod\limits_{i\in I} F_i$, if all $(F_i \colon \C \to \D)_{i\in I}$ do so,
    $\C$ has $I$-indexed coproducts, and $\D$ is $I$-extensive.
  \end{enumerate}
\end{propC}
\bknote{}%
\begin{rem}
  The original proof of \itemref{morEndoPrecise}{constantPrecise} in~\cite{WDKH2019} is inaccurate:
  the authors claimed that a morphism of the form $S \to F0$ is always $F$-precise, for any functor $F$.
  This is, however, not true in general: a simple counterexample is given by
  the category of pointed sets and $F = \Id$, since the square
  \[
    \begin{tikzcd}
      \set{\ast,x}
      \arrow{r}{}
      \arrow{d}[swap,pos=0.4]{\id}
      & \set{\ast}
      \arrow{d}{}
      \\
      \set{0,\ast}
      \arrow{r}[swap]{}
      & \set{\ast}
    \end{tikzcd}
  \]
  has no diagonal fill-in ($\ast$ is the basepoint). Fortunately, the claim does indeed hold
  if $F$ is a constant functor, so the proposition is still valid.
\end{rem}

\begin{cor}\label{corPolyPreciseFact}
Let $\C$ be an extensive category with finite products. Then all polynomial endofunctors on $\C$ admit precise factorizations \wrt $\Mor$.
\end{cor}
\begin{proof}
Consequence of \cref{idPreciseFact,morEndoPrecise}.
\end{proof}

\Cref{morEndoPrecise} provides sufficient conditions for a functor to admit precise factorizations only \wrt $\Mor$.
Fortunately, it is possible to extend this property to precise factorizations with respect to any factorization system:

\begin{prop}\label{preciseFactMorM}
  Suppose $F \colon \C \to \D$ admits precise factorizations \wrt $\Mor$.
  Then for any $(\E,\M)$-factorization system on $\C$, $F$ admits precise factorizations \wrt $\M$.
\end{prop}
\begin{proof}
  Let $f \colon P \to FY$ be a morphism, and take its precise factorization $(R,p,h)$ \wrt $\Mor$:
  \[
    \begin{tikzcd}
      P
      \arrow[dashed]{r}{p}
      \arrow{dr}[swap]{f}
      & FR
      \arrow[dashed]{d}{Fh}
      \\
      & FY
    \end{tikzcd}
  \]
  Moreover, take the $(\E,\M)$-factorization of the morphism $h \colon R \to Y$:
  \[
    \begin{tikzcd}
      R
      \arrow[->>]{r}{e}
      \arrow[shiftarr={yshift=6mm}]{rr}{h}
      & Q
      \arrow[>->]{r}{k}
      & Y
    \end{tikzcd}
  \]
  We claim that $r := Fe \cdot p$ is $F$-precise \wrt $\M$.
  Since $k \in \M$ and $Fk \cdot r = Fh \cdot p = f$,
  it then follows that $(Q, r, k)$ is the precise factorization of $f$ \wrt $\M$.

  It remains to show that $r$ is $F$-precise \wrt $\M$. Consider the commutative square
  below on the left with $m,n \in \M$:
  \[
    \begin{tikzcd}
      P
      \arrow{r}{p}
      \arrow{d}[swap]{g}
      \arrow[shiftarr={yshift=6mm}]{rr}{r}
      & FR
      \arrow{r}{Fe}
      & FQ
      \arrow{d}{Fm}
      \\
      FC
      \arrow{rr}[swap]{Fn}
      && FD
    \end{tikzcd}
    \quad\overset{\exists s}{\Longrightarrow}\quad
    \begin{tikzcd}
      P
      \arrow{r}{p}
      \arrow{d}[swap]{g}
      & FR
      \arrow{dl}{Fs}
      \\
      FC
    \end{tikzcd}
    \&\ 
    \begin{tikzcd}
      R
      \arrow[->>]{r}{e}
      \arrow{d}[swap]{s}
      & Q
      \arrow[>->]{d}{m}
      \arrow[dashed]{dl}[description]{d}
      \\
      C
      \arrow[>->]{r}[swap]{n}
      & D
    \end{tikzcd}
  \]
  Since $p$ is precise \wrt $\Mor$, we get a diagonal $s$ as shown above on the right.
  Furthermore, the commutative square $m \cdot e = n \cdot s$ with $e \in \E$ and $n \in \M$
  gives us a diagonal fill-in $d$, courtesy of the factorization system.
  Then $d$ is the required diagonal since $n \cdot d = m$ and
  \[ Fd \cdot r = Fd \cdot Fe \cdot p = Fs \cdot p = g. \qedhere \]
\end{proof}

\begin{exa}\label{bagPreciseFact}
  The bag functor~$\Bag$ admits precise factorizations.
  For the factorization of a function $f\colon X\to \Bag Y$,
  we create for each $x\in X$ and $y\in Y$ as many copies of $y$ as given by the
  multiplicity $f(x)(y) \in \N$.
  Let $Z = \coprod_{(x,y)\in X\times Y} \ceil{f(x)(y)}$.
  Then, define $m\colon Z\to Y$ by $m(\inj_{(x,y)}(k)) = y$ and $p\colon X\to \Bag Z$
  by $p(x)(\inj_{(x',y)}(k)) = 1$ if $x= x'$ and 0 otherwise.
\end{exa}

\Cref{bagPreciseFact} is a special case of the following result:
\begin{propC}[{\cite[Prop.~4.3]{WDKH2019}}]\label{analyticalPreciseFactor}
  Every analytical $\Set$-functor admits precise factorizations \wrt $\Mor$.
\end{propC}

On $\Set$, the \emph{finitary} analytical functors are precisely those functors that weakly preserve wide pullbacks~\cite[Cor.~2.7]{av08}.
For an arbitrary functor, we can show that (weak) preservation of (weak) wide $\M$-intersections
is closely related to the functor admitting precise factorizations:

\begin{thm}\label{thmPreciseIffWidePullback}
Let $\M$ be a class of morphisms in~$\C$.
If a functor $F \colon \C \to \D$ admits precise factorizations \wrt $\M$, then it preserves weak wide $\M$-intersections.
\twnote{}\bknote{}\bknote{}%
\end{thm}
\begin{proof}
  Assume that $F$ admits precise factorizations, and consider a family
  of $\M$-morphisms
  \[
    m_i\colon M_i\monoto Y
    \text{ in  }\M
    \qquad
    \text{for }i\in I
  \]
  that has a wide pullback
  \[
    g\colon C\to Y
    \text{ and }
    c_i\colon C\to
    M_i
    \text{ with }
    m_i\cdot c_i = g
    \qquad
    \text{ for }i\in I
  \]
  in $\C$.
  In order to verify that the cone $(FC, Fg, Fc_i)_{i\in I}$ is a weak wide pullback of $(Fm_i)_{i \in I}$ (in $\D$),
  consider another cone
  \[
    f\colon P\to FY
    \text{ and }
    d_i\colon P\to FM_i
    \text{ with }
    Fm_i\cdot d_i = f
    \qquad
    \text{ for }i\in I.
  \]
  We need to construct a cone morphism $(P, f, d_i)_{i \in I} \to (FC, Fg, Fc_i)_{i \in I}$. To this end, 
  take the precise factorization (\cref{defPreciseFactor}) of the morphism $f\colon P\to FY$:
  \[
    \begin{tikzcd}
      P
      \arrow{dr}[swap]{f}
      \arrow[dashed]{r}{p}
      & FR
      \arrow[dashed]{d}{Fh}
      \\
      &  FY
    \end{tikzcd}
    \qquad\text{ with }h\in \M
  \]
  Since $Fh\cdot p = f = Fm_i\cdot d_i$, the outside of the following square commutes
  for every $i\in I$:
  \[
    \begin{tikzcd}
      P
      \arrow{r}{p}
      \arrow{d}[swap]{d_i}
      & FR
      \arrow[dashed]{dl}[description]{Fr_i}
      \arrow{d}{Fh}
      \\
      FM_i
      \arrow{r}[swap]{Fm_i}
      & FY
    \end{tikzcd}
  \]
  Using that $h\in \M$ and $m_i\in \M$,
  the precise morphism $p$ induces a morphism $r_i\colon R\to M_i$ with
  $m_i\cdot r_i = h$ and $Fr_i\cdot p = d_i$, for all $i\in I$.
  In particular, $(R,h,(r_i)_{i\in I})$ forms a cone for the diagram of the
  $m_i$, and so the wide pullback $C$ induces a morphism $u \colon R \to C$ such that
  \[
    g \cdot u = h
    \text{ and }
    c_i \cdot u = r_i
    \qquad
    \text{ for every }i\in I.
  \]
  The composition $Fu\cdot p\colon P\to
  FC$ is then the desired cone morphism
  because the diagrams
  \[
    \begin{tikzcd}[sep=10mm]
      P
      \arrow{dr}[swap]{f}
      \arrow{r}{p}
      & FR
      \arrow{d}[pos=0.3]{Fh}
      \arrow{r}{Fu}
      & FC
      \arrow{dl}{Fg}
      \\
      & FY
    \end{tikzcd}
    \text{ and }
    \begin{tikzcd}[sep=10mm]
      P
      \arrow{dr}[swap]{d_i}
      \arrow{r}{p}
      & FR
      \arrow{d}[pos=0.3]{Fr_i}
      \arrow{r}{Fu}
      & FC
      \arrow{dl}{Fc_i}
      \\
      & FM_i
    \end{tikzcd}
  \]
  commute for every $i \in I$.
\end{proof}

\begin{cor}\label{setPreciseFactor}
  For a finitary functor $F\colon \Set\to\Set$, the following are equivalent:
  \begin{enumerate}
  \item\label{setPreciseFactorAna} $F$ is analytical.
  \item\label{setPreciseFactorPres} $F$ weakly preserves wide pullbacks.
  \item\label{setPreciseFactorFact} $F$ admits precise factorizations \wrt $\Mor$.
  \end{enumerate}
\end{cor}
\begin{proof}
  The equivalence of \labelcref*{setPreciseFactorAna} and \labelcref*{setPreciseFactorPres} is~\cite[Cor.~2.7]{av08}. 
  The implication \labelcref*{setPreciseFactorAna} $\Rightarrow$ \labelcref*{setPreciseFactorFact} is \cref{analyticalPreciseFactor},
  and the implication \labelcref*{setPreciseFactorFact} $\Rightarrow$ \labelcref*{setPreciseFactorPres} is (a consequence of) \cref{thmPreciseIffWidePullback}.
\end{proof}
\begin{rem}
  If $\M \subseteq \Mono$, $\C$ has wide $\M$-intersections, and $F$ weakly preserves them, then the converse of
  \cref{thmPreciseIffWidePullback} also holds~\cite[Prop.~5.9]{wmkd20reachability}.
  \jrnote{}\bknote{}
  Note that the cited proposition concerns least bounds instead of precise factorizations. However, under the given assumptions, the two notions are equivalent, as we shall prove next (\cref{preciseLeastCoincide}).
  \twnote{}\bknote{}%
  \bknote{}\bknote{}%
\end{rem}

\subsection{Comparison to least bounds}

If we instantiate $\M$ to be a class of monomorphisms, then the notion of precise factorizations
coincides with that of \emph{least bounds} (also called
\emph{base}~\cite{Blok12,BarloccoEA19}) for a functor:
\begin{defiC}[{\cite{Blok12,wmkd20reachability}}]
  For a functor $F\colon \C\to \D$ and a class $\M$ of monomorphisms in $\C$, we say
  that $F$ has \emph{least bounds (\wrt $\M$)} if for every morphism $f\colon
  X\to FY$, there is a least morphism $m\colon Z\monoto Y$ in $\M$ such that $f$
  factors through $Fm$ (via some $g\colon X\to FZ$). Concretely, this means
  that
  \[
    \begin{tikzcd}
      X
      \arrow{dr}[swap]{f}
      \arrow[dashed]{r}{g}
      & FZ
      \arrow[dashed]{d}{Fm}
      \\
      & FY
    \end{tikzcd}
  \]
  commutes and for every $m'\colon Z'\monoto Y$ and $g'\colon X\to FZ'$ with
  $Fm'\cdot g' = f$, there is a morphism $h\colon Z\to Z'$ with $m'\cdot h =
  m$.
  The triple $(Z,g,m)$ is called the \emph{least bound} of $f$.
\end{defiC}
In $\Set$, $Z$ intuitively consists of those elements of $Y$ that are actually used by $f$.
Then $m$ is just the subset inclusion, and since $m$ \textqt{omits} only those
elements which are not used by $f$ anyway, we can restrict the codomain of $f$
to $FZ$, yielding $g\colon X\to FZ$.

\begin{exa}\label{leastBoundFig3}
  For $\Set$ and $\M$ being the class of all monomorphisms (\ie injective maps), consider
  the map $f\colon X\to \FBinTree Y$ from \cref{figExPrecise} (where $\FBinTree$ is as in \itemref{ex:coalgebras}{ex:bintree}).
  The least bound of the map is given by
  \[
    Z = \set{y_1,y_2},
    \quad
    m \colon Z\hookrightarrow Y,
    \quad
    g\colon X\to \FBinTree Z,
    \quad
    g(x) = f(x).
  \]
  Here, $m$ is the subset inclusion and $g$ witnesses that the codomain of $f$ restricts to $\FBinTree Z$.
\end{exa}
\begin{exa}
  The bag functor $\Bag$ on sets has least bounds \wrt $\Mono$. The least
  bound of $f\colon X\to \Bag Y$ is given by those elements of $Y$ that appear
  with non-zero multiplicity: $Z := \set{y\in Y\mid \exists x\in X\colon
  f(x)(y) \neq 0}$ and $f$ then factors through $\Bag m$ for $m\colon
  Z\hookrightarrow Y$.
\end{exa}

A key observation for the generalization of reachability to possibly
non-proper factorization systems in \cref{sec:reach} is that
precise factorizations instantiate to least bounds in the following sense:

\begin{prop}
\label{preciseLeastCoincide}
  Let $F \colon \C \to \D$, let $\M \subseteq \Mono$, and let $f \colon X \to FY$. Then:
  \begin{enumerate}
  \item\label{preciseImpliesLeast} If the precise factorization (\wrt $\M$) of $f$ exists, then
  it is also the least bound of $f$.

  \item\label{leastImpliesPrecise} If $\C$ has and $F$ weakly preserves $\M$-intersections,
  and if the least bound of $f$ exists,
  then the least bound of $f$ is also the precise factorization of $f$.
  \end{enumerate}
  In particular, under the given conditions, each of them exists iff the other one does.
\end{prop}
\begin{proof}\hfill
\begin{enumerate}
\item 
  Consider the precise factorization of $f\colon X\to FY$:
  \[
    \begin{tikzcd}
      X
      \arrow{dr}[swap]{f}
      \arrow[dashed]{r}{g}
      & FZ
      \arrow[dashed]{d}{Fm}
      \\
      & FY
    \end{tikzcd}
  \]
  For the verification that $(Z,g,m)$ is the least bound of $f\colon X\to FY$, consider
  $(Z',g',m')$ with $m'\in\M$ and
  \[
    \begin{tikzcd}
      X
      \arrow{dr}[swap]{f}
      \arrow{r}{g'}
      & FZ'
      \arrow{d}{Fm'}
      \\
      & FY
    \end{tikzcd}
  \]
  Thus, the following square commutes.
  \[
    \begin{tikzcd}
      X
      \arrow{r}{g}
      \arrow{d}[swap]{g'}
      & FZ
      \arrow{d}{Fm}
      \arrow[dashed]{dl}[description]{Fh}
      \\
      FZ'
      \arrow{r}[swap]{Fm'}
      & FY
    \end{tikzcd}
  \]
  Since $m,m'\in \M$, the precise morphism $g$ induces $h\colon Z\to Z'$
  satisfying $m'\cdot h = m$, as desired.

\item
  Consider the least bound $(Z,g,m)$ of $f\colon X\to FY$:
  \[
    \begin{tikzcd}
      X
      \arrow{dr}[swap]{f}
      \arrow[dashed]{r}{g}
      & FZ
      \arrow[dashed]{d}{Fm}
      \\
      & FY
    \end{tikzcd}
  \]
  For the verification that $g\colon X\to FZ$ is precise, we use
  \cref{preciseWithIntersections} and consider morphisms
  $u \colon X \to FW$ and $w \colon W \monoto Z$ with $w \in \M$ and $Fw \cdot u = g$.
  Then we have $F(m\cdot w)\cdot u = Fm \cdot g = f$.
  Thus, the minimality of $m$ induces a morphism $h\colon Z\to W$ with
  $(m\cdot w) \cdot h = m$. Since $m\in \M\subseteq \Mono$, this shows $w\cdot h = \id_Z$.
  Furthermore, we have $Fw \cdot Fh \cdot g = F\id_Z \cdot g = Fw \cdot u$.
  Since $Fw \in \Mono$ by \cref{lemIntsecMonoPres}, we acquire the other desired equality $Fh \cdot g = u$.
  \qedhere
\end{enumerate}
\end{proof}

\begin{exa}
Consider the least bound~$(Z, g, m)$ of~$f$ from \cref{leastBoundFig3}.
Since the functor~$\FBinTree$ (\itemref{ex:coalgebras}{ex:bintree}) preserves intersections and monos,
\itemref{preciseLeastCoincide}{leastImpliesPrecise} implies
that $(Z, g, m)$ is also the precise factorization of~$f$ \wrt $\Mono$. Thus, the map~$g$ is precise
\wrt $\Mono$. However, a straightforward adaption of \cref{exPreciseMap} (replacing $Y$ with $Z$) shows that $g$ is not precise \wrt $\Mor$.
\end{exa}

Combining our results with those of~\cite{wmkd20reachability}, we arrive at the following \lcnamecref{admitPreciseIffHasLeast}:
\begin{cor}\label{admitPreciseIffHasLeast}
  Let $\M \subseteq \Mono$ and suppose that $\C$ has $\M$-intersections.
  Then a functor $F \colon \C \to \D$ admits precise factorizations \wrt $\M$ iff it has least bounds \wrt $\M$ and sends $\M$-morphisms to monos.
  Furthermore, if any of these equivalent conditions holds, then precise factorizations coincide with least bounds.
\end{cor}
\begin{proof}
  If $F$ admits precise factorizations, then, by \itemref{preciseLeastCoincide}{preciseImpliesLeast},
  the least bound of any morphism exists and coincides with its precise factorization.
  Furthermore, by \cref{thmPreciseIffWidePullback}, $F$ preserves weak $\M$-intersections,
  so \cref{lemIntsecMonoPres} implies that it sends $\M$-morphisms to monos.
  Conversely, assume that $F$ has least bounds and sends $\M$-morphisms to monos.
  Then, by~\cite[Prop.~5.9]{wmkd20reachability}, it preserves $\M$-intersections.
  Thus, \itemref{preciseLeastCoincide}{leastImpliesPrecise} implies that
  the least bound of any morphism is also its precise factorization.
\end{proof}

\section{Generalized Minimality}
\label{sec:genmin}

In a previous paper~\cite{WKRT25}, we extended both the definition and the reachability
construction to work for general $(\E,\M)$-factorization systems, even those
that are not proper. In the present article, we connect these definitions of reachability
in a category of coalgebras to $\M$-minimality in a category studied in earlier work~\cite{Wissmann22},
thereby obtaining a joint generalization of both concepts.
Choosing our base category to be a category of coalgebras, we recover
our original generalized reachability notion; this is done in \cref{sec:reach}.
If, in addition, we let $\M$ be the class of all morphisms,
the notion of minimality instantiates
to being tree-shaped, as discussed in \cref{sec:trees}.
\twnote{}
\rtnote{}\bknote{}%
\twnote{}
\rtnote{}

\begin{defi}\label{defMinimal}
  Let $\C$ be a category, and fix a class~$\M$ of morphisms in~$\C$.
  \begin{enumerate}
  \item An object~$C\in \C$ is \emph{$\M$-(split-)minimal} if every
  morphism $m\colon T\monoto C$ in~$\M$ is a split epimorphism.
  \item The \emph{$\M$-minimization} of an object~$D\in \C$ is an $\M$-morphism
  $h\colon C\monoto D$ where $C$ is $\M$-split-minimal.
  \end{enumerate}
\end{defi}

\noindent For historical comparison, we refer to the earlier minimality
notion as \emph{$\M$-iso-minimality}. The unqualified term
\emph{$\M$-minimal} will always mean $\M$-split-minimal.
\begin{defiC}[{\cite[Def.~4.1]{Wissmann22}}]
  An object~$C\in \C$ is \emph{$\M$-iso-minimal} if every 
  morphism $m\colon T\monoto C$ in~$\M$ is an isomorphism.
\end{defiC}

\begin{lem}\label{lemMonoMinimal}
  If $\M\subseteq \Mono$, then $\M$-split-minimality and $\M$-iso-minimality
  coincide.
\end{lem}
\begin{proof}
  Since every isomorphism is a split epimorphism,
  every $\M$-iso-minimal object is also $\M$-split-minimal.
  Conversely, if $C$ is $\M$-split-minimal, then every morphism
  $m\colon T\monoto C$ in~$\M$ is both a split epimorphism
  and a monomorphism, and thus an isomorphism.
\end{proof}

We now investigate some properties of $\M$-minimality.
When $\M \subseteq \Mono$, all results of~\cite{Wissmann22} apply
by \cref{lemMonoMinimal}. Some results also hold under weaker assumptions.
For instance, if $\C$ has an $(\E,\M)$-factorization system,
then the following characterization of $\M$-minimality in terms of~$\E$ is immediate:
\begin{lemC}[{\cite[Lem.~4.6]{Wissmann22}}]\label{minEveryE}
  If $\M$ is part of an $(\E,\M)$-factorization system and $\SplitEpi \subseteq \E$,
  then an object~$C$ is $\M$-minimal iff every morphism $h\colon D\to C$
  is in~$\E$.
\end{lemC}
\begin{proof}
  ($\Leftarrow$) Consider an $\M$-morphism $h\colon D\to C$.
  Since $h\in \E$ by assumption, we have $h\in \M\cap\E = \Iso$.
  In particular, $h$ is a split epimorphism.

  ($\Rightarrow$):
  Let $h \colon D \to C$ be a morphism, and take its $(\E,\M)$-factorization:
  \[
      \begin{tikzcd}[baseline=(A.base),row sep=2mm,column sep=4mm]
        |[alias=A]|
        D
        \arrow[->>,to path={|- (\tikztotarget) \tikztonodes},rounded corners]{dr}[swap,pos=0.3]{e}
        \arrow[]{rr}[alias=f]{h}
        && C
        \\
        & \Im (h)
        \arrow[>->,to path={-| (\tikztotarget) \tikztonodes},rounded corners]{ur}[swap,pos=0.7]{m}
      \end{tikzcd}
  \]
  Because $C$ is $\M$-minimal and $m \in \M$, $m$ is a split epimorphism.
  Since $\SplitEpi \subseteq \E$, we also have $m \in \E$.
  Hence, $h = m \cdot e \in \E$.
\end{proof}
\begin{rem}
  For an $(\E,\M)$-factorization system, the condition $\M \subseteq \Mono$ always
  implies $\SplitEpi \subseteq \Mono$ by the dual of~\cite[Prop.~14.10(1)]{joyofcats}.
  Furthermore, if $\C$ has binary products, then the converse also holds
  by the dual of~\cite[Prop.~14.11]{joyofcats}.
\end{rem}

In the following \lcnamecref{minimalUnique}, we say that a cospan
$B \xrightarrow{f} D \xleftarrow{g} C$ can be \emph{closed}
if there exist morphisms $p$~and~$q$ that complete the cospan
into a commutative square:
\[
  \begin{tikzcd}[sep=6mm]
    A
    \arrow[dashed]{d}[swap]{p}
    \arrow[dashed]{r}{q}
    & C
    \arrow{d}{g}
    \\
    B
    \arrow{r}[swap]{f}
    & D
  \end{tikzcd}
\]
\begin{lem}\label{minimalUnique}
  Let $\M$ be part of an $(\E,\M)$-factorization system.
  \begin{enumerate}
  \item If $h\colon C\monoto D$ is an $\M$-morphism between $\M$-minimal
  objects, then $h$ is an isomorphism.
  \item\label{minimalUniqueStm} If all cospans in~$\C$ can be closed, then any two minimizations
  of an object~$D \in \C$ are isomorphic as objects of the slice category~$\C/D$.
  \rtnote{}\bknote{}
  \end{enumerate}
\end{lem}
\noindent Thus, we may speak of \emph{the} $\M$-minimization of objects (if it exists).
\begin{proof}\hfill
\begin{enumerate}
\item For $h\colon C\monoto D$ in~$\M$, the $\M$-minimality of~$D$ induces some
  $s\colon D\to C$ with $h\cdot s = \id_D$.
  Given that $h\in \M$ and $\id_D\in \M$, the
  cancellation property (\cref{remCancel}) yields $s\in \M$.
  Since $C$ is also $\M$-minimal, we obtain $t\colon C\to D$ with
  $s\cdot t = \id_C$. Thus, $s$ has both a retraction~$h$ and a section~$t$,
  implying that $h$ is an isomorphism.

\item
  Consider $D\in \C$ with two $\M$-minimizations $g \colon R \monoto D$ and $h \colon S \monoto D$.
  By assumption, we can close the cospan into a commutative square:
  \[
    \begin{tikzcd}
      & D
      \\
      R
      \arrow[>->]{ru}{g}
      && S
      \arrow[>->]{ul}[swap]{h}
      \\
      & P
      \arrow[dashed]{ul}{q}
      \arrow[dashed]{ur}[swap]{p}
    \end{tikzcd}
  \]
  Take the $(\E,\M)$-factorization $r\cdot e$ of $q$:
  \[
    \begin{tikzcd}
      & D
      \\
      R
      \arrow[>->]{ru}{g}
      \arrow[dashed,>->,bend left=20,shift=({0.5mm,0.5mm})]{dr}{f}
      && S
      \arrow[>->]{ul}[swap]{h}
      \\
      & T
      \arrow[>->]{ul}{r}
      \arrow[dashed,>->]{ur}{s}
      &
      \\
      & P
      \arrow[->>]{u}[pos=0.4]{e}
      \arrow[bend left=15]{uul}{q}
      \arrow[bend right=15]{uur}[swap]{p}
    \end{tikzcd}
  \]
  Since $h\in \M$ and $e \in \E$, the factorization system gives us a diagonal fill-in $s\colon T\to S$.
  As $h \cdot s = g \cdot r$ and $h, g \cdot r \in \M$, $s$ is necessarily in~$\M$
  by the cancellation property (\cref{remCancel}).
  By the $\M$-minimality of~$R$, we obtain a splitting of~$r$, that is, a
  morphism $f\colon R\to T$ with $r\cdot f = \id_R$.
  Using the cancellation property again, $r \in \M$ and $\id_R \in \M$ imply $f\in \M$.
  Then the composed morphism $s\cdot f\colon R\to S$ between two
  $\M$-minimal objects is in~$\M$, and thus an isomorphism by the previous item.
  Furthermore, we have $h \cdot s \cdot f = g \cdot r \cdot f = g$,
  so $s \cdot f$ is an isomorphism $g \cong h$ in~$\C/D$.
  \qedhere
\end{enumerate}
\end{proof}

\noindent There is a strong connection between the notion of $\M$-minimality and $\M$-projectivity.
Given a class~$\M$ of morphisms, an object~$A \in \C$ is called \emph{$\M$-projective} if
for all morphisms $h\colon A\to C$ and $m\colon B\monoto C$ in~$\M$,
there is some morphism $u\colon A\to B$ such that the diagram below commutes:
\[
  \begin{tikzcd}
    & B
    \arrow[>->]{d}{m}
    \\
    A
    \arrow{r}[swap]{h}
    \arrow[dashed]{ur}{u}
    & C
  \end{tikzcd}
\]

\begin{thm}\label{thm:projective}
  Every $\M$-projective object is $\M$-minimal. Moreover, if the class $\M$ is part of an $(\E, \M)$-factorization system and all cospans in~$\C$ can be closed, then every $\M$-minimal object is $\M$-projective.
\end{thm}
\noindent Note that the above induced morphism $u$ is not required to be unique.
For example, tree unravellings of graphs are not unique up to \emph{unique}
isomorphism (because isomorphic subtrees can be permuted). Likewise, the induced morphism $u$ is not unique in general.
A concrete example where $u$ is not unique is discussed later in \cref{exWhyNotUniqu}.
\twnote{}
\bknote{}%
\begin{proof}
  ($\Rightarrow$) Let $D$ be $\M$-projective, and consider an $\M$-morphism
  $m\colon C\monoto D$. Using the assumed projectivity property with $h := \id_D$,
  we obtain a morphism $s \colon D \to C$ completing the triangle:
  \[
    \begin{tikzcd}
      & C
      \arrow[>->]{d}{m}
      \\
      D
      \arrow{r}[swap,pos=0.6]{\id_M}
      \arrow[dashed]{ur}{s}
      & D
    \end{tikzcd}
  \]
  This verifies that $m$ is a split epimorphism, showing that $D$ is $\M$-minimal,
  as desired.
  
  ($\Leftarrow$) 
  Suppose $M$ is $\M$-minimal, and let $m \colon D \monoto C$ in~$\M$ and $h \colon M \to C$.
  By assumption, we can close this cospan into a commutative square:
  \[
    \begin{tikzcd}
      P
      \arrow[dashed]{r}{p}
      \arrow[dashed]{d}[swap]{q}
      & D
      \arrow[>->]{d}{m}
      \\
      M
      \arrow{r}[swap]{h}
      & C
    \end{tikzcd}
  \]
  Consider the $(\E, \M)$-factorization $n \cdot e$ of $q$:
  \[
    \begin{tikzcd}
      P
      \arrow{r}{p}
      \arrow[->>]{d}[swap,pos=0.4]{e}
      & D
      \arrow[>->]{dd}{m}
      \\
      Q
      \arrow[>->]{d}{n}
      \arrow[dashed]{ur}[swap,pos=0.4]{d}
      \\
      M
      \arrow{r}[swap]{h}
      \arrow[dashed,bend left=20,xshift=-0.5mm]{u}{s}
      & C
    \end{tikzcd}
  \]
  Since $M$ is $\M$-minimal, $n$ has a splitting $s\colon M\to Q$ with $n\cdot s = \id_M$.
  Moreover, given that $e \in \E$ and $m \in \M$, we get a diagonal fill-in $d\colon Q\to D$ such that
  $d \cdot e = p$ and $m \cdot d = h \cdot n$.
  Then $u:= d\cdot s$ provides the desired lift since
  \[ m \cdot u = m \cdot d \cdot s = h \cdot n \cdot s = h. \qedhere \]
\end{proof}

\section{Generalized Reachability}
\label{sec:reach}

When now instantiating $\M$-minimality in a category of pointed coalgebras, we obtain the notion of \emph{generalized reachability} introduced in~\cite{WKRT25}.

\subsection{Universal Property}
The universal property of reachability captures that it is not possible to
\textqt{omit} states. The omission of states is formally encoded by a coalgebra
morphism from a class $\M$:
\begin{defi}\label{defReachable}
  Fix a functor $F\colon \C\to \C$ and a class~$\M$ of morphisms in~$\C$.
  \begin{enumerate}
  \item A \emph{pointed $\M$-subcoalgebra} of a pointed coalgebra~$(C,c,i_C)$
  is an $\liftfactsys{\M}{F}$-morphism $(T,t,i_T)\to (C,c,i_C)$.
  We simply speak of \emph{subcoalgebras} if $\M$ is clear from the context
  (even if $\M$ is not assumed to be a class of monomorphisms).
  \item\label{itemMReachable} A coalgebra~$(C,c,i_C)$ is called
  \emph{($\M$-)reachable} if each pointed $\M$-subcoalgebra of~$(C,c,i_C)$
  is a split epimorphism of coalgebras
  (\ie the splitting is also a pointed homomorphism).
  \end{enumerate}
\end{defi}
\noindent Note that the term `subcoalgebra' needs to be understood abstractly
because it is parametric in the class~$\M$, which may contain morphisms that are
not monic. Still, we stick to this terminology because it conveys the
right intuition if $\M\subseteq \Mono$ is assumed.
Under this latter assumption, we recover the usual definition of reachability~\cite{AMMS03}:
\begin{lem}
  \label{lemMonoReachable}
  If $\M\subseteq \Mono$, then a pointed $F$-coalgebra is reachable (\cref{defReachable}) iff every one of its pointed subcoalgebras is an isomorphism.
\end{lem}
\begin{proof}
  Special case of \cref{lemMonoMinimal} applied to~$\C := \Coalg[I](F)$
  and~$\M := \liftfactsys{\M}{F}$.
  Note that $\liftfactsys{\M}{F}$-split-minimality in~$\Coalg[I](F)$
  coincides with $\M$-reachability,
  $\liftfactsys{\M}{F}$-iso-minimality in~$\Coalg[I](F)$
  coincides with the condition on the right hand side,
  and if $\M \subseteq \Mono$ in~$\C$,
  then $\liftfactsys{\M}{F} \subseteq \Mono$ in~$\Coalg[I](F)$.
\end{proof}
\begin{exa}\label{bagReachable}
  For $\M = \Mono$, a directed graph viewed as a $\Pow$-coalgebra
  is reachable iff it is reachable in the usual
  graph-theoretic sense~\cite[Ex.~2.7(1)]{wmkd20reachability}.
  Similarly, a pointed directed multigraph is $\Mono$-reachable
  as a coalgebra~$(V,c,v_0)$ iff for every vertex~$u\in V$,
  there is some path from~$v_0$ to~$u$, that is, iff
  $\Path(v_0,u)\neq \emptyset$ for all~$u\in V$.
\end{exa}
\begin{exa}\label{dfaReachable}
  For $\M=\Mono$, a partial deterministic automaton $\fpair{o,\delta}\colon C\to 2\times
  (C+\set{\bot})^A$, considered as an $\FAut{A}$-coalgebra (\itemref{ex:coalgebras}{ex:pdfa}) and equipped with a point $q_0\colon 1\to C$,
  is reachable iff for each state $q\in C$,
  there is an input word $w\in A^*$ such that $\delta^*(q_0)(w) = \inl(q)$.
\end{exa}

\begin{figure}
    \begin{tikzpicture}[x=12mm]
      \begin{scope}[coalgebra,scope of math nodes,x=10mm,local bounding box=CoalgC]
        \node[initial,state] (q0) at (0,0) {q_0};
        \node[state] (q1) at (1,0) {q_1};
        \node[state] (q2) at (2,0) {q_2};
        \node (dots) at (2.8,0) {\cdots};
        \path[transition] (q0) to (q1);
        \path[transition] (q1) to (q2);
        \path[transition,-] (q2) to (dots);
      \end{scope}
      \begin{scope}[coalgebra,scope of math nodes,xshift=6.0cm,local bounding box=CoalgD]
        \node[initial,state] (p0) at (0,0) {p_0};
        \node[state] (p1) at (1,0) {p_1};
        \path[transition,bend right=15] (p0) to (p1);
        \path[transition,bend right=15] (p1) to (p0);
      \end{scope}
      \begin{scope}[coalgebra,scope of math nodes,xshift=10.1cm,local bounding box=CoalgE]
        \node[initial,state] (l) at (0,0) {\ell};
        \path[transition,loop at = 90,looseness=4,overlay] (l) to (l);
      \end{scope}
      \node[coalgname,outer sep=0pt,anchor=south west] (C label) at (CoalgC.north west) {\ensuremath{C}};
      \node[coalgname,outer sep=0pt,anchor=south west] (D label) at (CoalgD.north west) {\ensuremath{D}};
      \node[coalgname,outer sep=0pt,anchor=south west] (E label) at ([xshift=-2mm]CoalgE.north west) {\ensuremath{E}};
      \begin{scope}[on background layer]
        \node[coalgebra frame,fit=(CoalgC) (C label)] (FrameC){};
        \node[coalgebra frame,fit=(CoalgD) (D label)] (FrameD){};
        \node[coalgebra frame,fit=(CoalgE) (E label)] (FrameE){};
      \end{scope}
      \begin{scope}[commutative diagrams/.cd,every diagram,
                   ]
        \path[commutative diagrams/every arrow,
                   shorten >= 1mm,
                   shorten <= 1mm] (FrameC)
          to node[commutative diagrams/every label] {\ensuremath{h}} (FrameD);
        \path[commutative diagrams/every arrow,
                   shorten >= 1mm,
                   shorten <= 1mm] (FrameD.east)
          to node[commutative diagrams/every label] {\ensuremath{g}} (FrameE.west |- FrameD.east);
      \end{scope}
    \end{tikzpicture}%
    \caption{Surjective coalgebra morphisms without splittings.}
    \label{figFoldLoop}
\end{figure}

\begin{exa}
  Let us briefly look at a
  small example to discuss split epimorphisms of coalgebras. Consider
  the $(\Iso,\Mor)$-factorization system in $\Set$ and coalgebras for the identity
  functor, visualized in \cref{figFoldLoop}.
  The map $g\colon (D,d,p_0) \to (E,e,\ell)$ is in $\Mor$. However,
  there is no coalgebra morphism $(E,e) \to (D,d)$ (regardless of point
  preservation), so $g$ is not a split epimorphism of pointed coalgebras.
  And indeed, $E$ is not a tree (by intuitive means), because it has a cycle.
  The underlying map $g\colon D\to E$ has two splittings ($s_1\colon
  \ell\mapsto p_0$ and $s_2\colon \ell\mapsto p_1$), showcasing why we require
  split epis of \emph{coalgebras} in \itemref{defReachable}{itemMReachable}.
  Likewise, $D$ is also not a tree (by intuitive means), because it has a cycle
  of length 2. And again, $D$ is not $\Mor$-reachable as witnessed by the
  coalgebra morphism $h$, which has no splitting in coalgebras (though being
  surjective).
  This example can be naturally adapted for coalgebras for polynomial functors
  with at least one symbol of arity at least 1.
  
  Related to this, in \cref{exWhySplit} later, we will see a concrete example
  in which the morphism induced by \itemref{defReachable}{itemMReachable} is
  only a split epimorphism instead of an isomorphism.
\end{exa}
\jrnote{}\bknote{}

\begin{lem}\label{reachableUnique}
  Let $\M$ be part of an $(\E,\M)$-factorization system.
  \begin{enumerate}
  \item\label{isoBetweenReachable}
        If $h\colon (C,c,i_C)\monoto (D,d,i_D)$ is an $\M$-morphism between $\M$-reachable coalgebras, then $h$ is an isomorphism.
  \item\label{reachableUniqueStm} If $\C$ has $\M$-intersections and $F$ preserves them weakly, then
        every coalgebra has at most one $\M$-reachable subcoalgebra.
  \end{enumerate}
\end{lem}
\noindent Thus, we may speak of \emph{the} $\M$-reachable subcoalgebra of a coalgebra (if it exists).
\begin{proof}
  The proof is a straightforward adaptation of the proof of \cref{minimalUnique}
  for the special case $\C := \Coalg[I](F)$ and $\M := \liftfactsys{\M}{F}$.
  For \labelcref*{reachableUniqueStm}, the conditions guarantee that
  $\Coalg[I](F)$ has $\M$-intersections (see \cref{exCoalgIntersection}),
  which implies that all cospans in~$\Coalg[I](F)$ can be closed.
  Furthermore, since $\M$-morphisms are stable under pullback (\cref{remCancel}),
  the morphism~$q$ in the proof of \itemref{minimalUnique}{minimalUniqueStm}
  may be assumed to be in~$\M$, so we can avoid any factorization in~$\Coalg[I](F)$
  (for which we would need the additional assumption that $F$ preserves $\M$).
\end{proof}
\begin{rem}
  In \itemref{reachableUnique}{isoBetweenReachable}, it is crucial that $h$ is
  required to be an $\M$-morphism. Otherwise, the statement is wrong
  for \eg $\M=\Mono$ and functors that do not preserve preimages. A concrete
  example is the functor $FX =
  \R^{(X)}$~\cite{Gumm2005}, for which successors in a coalgebra can disappear when
  being identified.
  \Cref{figNoPreim} shows examples of coalgebra morphisms for this functor.
  The composition $g\cdot h\colon C\to E$ is a morphism between reachable
  $F$-coalgebras, but $g\cdot h$ is not an isomorphism.
\end{rem}
\begin{figure}
  \begin{tikzpicture}[x=8mm,y=10mm,
      coalgebra/.append style={
        state/.append style={
          minimum size=4pt,
        },
      },
      ]
    \begin{scope}[coalgebra,local bounding box=CoalgC]
      \node[initial above,state] (p0) at (0,0) {};
      \node[state] (p1) at (0.7,-1) {};
      \node[state] (p2) at (-0.7,-1) {};
      \path[transition] (p0) to node[fill=coalgebraBackground,inner sep=1pt,minimum size=3mm] {\footnotesize 1} (p1);
      \path[transition] (p0) to node[fill=coalgebraBackground,inner sep=1pt,minimum size=3mm] {\footnotesize -1} (p2);
    \end{scope}
    \begin{scope}[coalgebra,xshift=30mm,local bounding box=CoalgD]
      \node[initial above,state] (q0) at (0,0) {};
      \node[state] (q1) at (0,-1) {};
    \end{scope}
    \begin{scope}[coalgebra,xshift=50mm,local bounding box=CoalgE]
      \node[initial above,state] (l) at (0,-0.5) {};
      \node[draw=none] at (0,-1) {};
    \end{scope}
    \node[coalgname,outer sep=0pt,anchor=north west] (C label) at ([yshift=0mm]CoalgC.north west) {\ensuremath{C}};
    \node[coalgname,outer sep=0pt,anchor=north west] (D label) at ([yshift=0mm,xshift=-5mm]CoalgD.north west) {\ensuremath{D}};
    \node[coalgname,outer sep=0pt,anchor=north west] (E label) at (D label.north -| CoalgE.west) {\ensuremath{E}};
    \begin{scope}[on background layer]
      \node[coalgebra frame,fit=(CoalgC) (C label)] (FrameC){};
      \node[coalgebra frame,fit=(CoalgD) (D label)] (FrameD){};
      \node[coalgebra frame,fit=(CoalgE) (E label)] (FrameE){};
    \end{scope}
    \begin{scope}[commutative diagrams/.cd,every diagram,
                  ]
      \path[commutative diagrams/every arrow,
                  shorten >= 1mm,
                  shorten <= 1mm] ([yshift=-3mm]FrameC.east)
        to node[commutative diagrams/every label] {\ensuremath{h}} ([yshift=-3mm]FrameD.west);
      \path[commutative diagrams/every arrow,
                  shorten >= 1mm,
                  shorten <= 1mm,
                rounded corners=2mm] ([yshift=-3mm]FrameD.east)
        -- node[commutative diagrams/every label,pos=0.5,above] {\ensuremath{g}} ([yshift=-3mm]FrameD.east -| FrameE.west);
    \end{scope}
  \end{tikzpicture}
  \caption{$\mathbb{R}^{(-)}$-coalgebra morphisms $g$ and $h$}
  \label{figNoPreim}
\end{figure}

\begin{lem}\label{reachEveryE}
  If $\M\subseteq\Mono$ is part of an $(\E,\M)$-factorization system and
  $F$ preserves $\M$, then a coalgebra~$C$ is $\M$-reachable iff
  every homomorphism $h\colon D\to C$ is in~$\E$.
\end{lem}
\begin{proof}
  Special case of \cref{minEveryE} for $\C := \Coalg[I](F)$ and the
  $(\liftfactsys{\E}{F},\liftfactsys{\M}{F})$-factorization system,
  which exists by \cref{coalgFact}.
\end{proof}

\subsection{Construction}\label{sec:reachablePartCons}
Having shown uniqueness, we now discuss sufficient conditions for
the existence of the $\M$-reachable subcoalgebra of a given pointed coalgebra. We do so
by generalizing the iterative reachability
construction~\cite{BarloccoEA19,wmkd20reachability} to arbitrary factorization
systems:

\begin{asm}\label{reachablePartAssumptions}
  For the remainder of Section~\thesubsection, we
  fix a functor $F\colon \C\to \C$ on a category~$\C$ with an
  $(\E,\M)$-factorization system such that $F$ preserves $\M$-morphisms
  and $F$ admits precise factorizations \wrt $\M$.
\end{asm}
\begin{cons}\label{levelConstruction}
  For a pointed $F$-coalgebra~$(C,c,i_C)$, we define a sequence
  of morphisms $m_k\colon C_k\monoto C$ in $\M$ and
  precise morphisms $c_k\colon C_k\to F C_{k+1}$ ($k\in \N$) inductively:
  \begin{enumerate}
  \item Take the $(\E,\M)$-factorization of $i_C\colon I \to C$:
    \begin{equation}\label{eq:constr:base}
      \begin{tikzcd}
        I
        \arrow[->>]{r}{i_C'}
        \arrow[shiftarr={yshift=6mm}]{rr}{i_C}
        & C_0
        \arrow[>->]{r}{m_0}
        & C
      \end{tikzcd}
    \end{equation}

  \item Given $m_k\colon C_k\to C$, let $c_{k}$ and $m_{k+1}\in \M$ be the precise
    factorization of~$c\cdot m_k$:
    \begin{equation}
      \begin{tikzcd}
        C_k
        \arrow[dashed]{r}{c_k}
        \arrow[>->,swap]{d}{m_k}
        & FC_{k+1}
        \arrow[dashed,>->]{d}[overlay]{Fm_{k+1}}
        \\
        C \arrow{r}[swap]{c}
        & FC
      \end{tikzcd}
      \label{eq:constr:step}
    \end{equation}
  \end{enumerate}
  We call $C_k$ the $k$-th level.
\end{cons}
\begin{lem}\label{coprodLevelsHom}
  In the setting of \cref{levelConstruction}, the coproduct of levels is a coalgebra making $[m_k]_{k\in \N}$ a pointed
  coalgebra morphism:
  \[
    \begin{tikzcd}
      I
      \arrow{r}{i_C'}
      \arrow{drr}[swap]{i_C}
      &
      C_0
      \arrow{r}{\inj_0}
      &
      \coprod_{k\in \N} C_k
      \arrow{r}[pos=0.45]{\coprod_{k\in \N}c_k}
      \arrow{d}{[m_{k}]_{k\in \N}}
      &[3mm] \coprod_{k \in \N} FC_{k+1}
      \arrow{r}{[F\inj_{k+1}]_{k\in \N}}
      &[6mm] F \coprod_{k\in \N} C_k
      \arrow{d}{F[m_{k}]_{k\in \N}}
      \\
      &
      &
      C
      \arrow{rr}[swap]{c}
      & & FC
    \end{tikzcd}
  \]
\end{lem}
\begin{proof}
  The preservation of the point follows immediately from the definition of~$m_0$~\labelcref{eq:constr:base}.
  For every~$\ell\in \N$, the outside of the following diagram commutes by the
  definition of~$c_k$~\labelcref{eq:constr:step}:
  \[
    \begin{tikzcd}[row sep=8mm, column sep=15mm]
      C_\ell
      \arrow{d}{\inj_{\ell}}
      \arrow[shiftarr={xshift=-12mm},>->]{dd}[swap]{m_{\ell}}
      \arrow{rr}{c_\ell}
      & & FC_{\ell+1}
      \arrow[shiftarr={xshift=12mm},>->]{dd}{Fm_{\ell+1}}
      \arrow{d}[swap]{F\inj_{\ell+1}}
      \\
      \coprod_{k\in \N} C_k
      \arrow{r}[pos=0.45]{\coprod_{k\in \N}c_k}
      \arrow{d}{[m_{k}]_{k\in \N}}
      & \coprod_{k \in \N} FC_k
      \arrow{r}{[F\inj_{k+1}]_{k\in \N}}
      \descto[pos=0.4]{d}{?}
      & F \coprod_{k\in \N} C_k
      \arrow{d}[swap]{F[m_{k}]_{k\in \N}}
      \\
      C
      \arrow{rr}[swap]{c}
      & {}
      & FC
    \end{tikzcd}
  \]
  Except for the cell marked \textqt{?}, all cells commute by basic properties of coproducts.
  Since the coproduct injections~$(\inj_{\ell})_{\ell\in \N}$ are jointly epic,
  the desired square \textqt{?} commutes as well.
\end{proof}
Taking the factorization of this homomorphism (in the category of pointed coalgebras) yields a pointed coalgebra again,
the reachable subcoalgebra of~$C$:
\begin{thm}\label{thm:reach}
  In the setting of \cref{levelConstruction}, the image~$R$ of the
  pointed coalgebra morphism~$[m_k]_{k\in \N}$ is an $\M$-reachable coalgebra:
  \[
      \begin{tikzcd}[column sep=12mm]
        \coprod_{k\in \N} C_k
        \arrow[->>]{r}[pos=0.45]{[e_k]_{k\in \N}}
        \arrow[shiftarr={yshift=8mm}]{rr}{[m_k]_{k\in \N}}
        & R
        \arrow[>->]{r}{m'}
        & C
      \end{tikzcd}
  \]
\end{thm}
\begin{rem}
  A restricted variant of this result was shown in~\cite[Theorem 17]{BarloccoEA19} with
  a fixed factorization system (where $\E$ is the class of strong epis and $\M=\Mono$, arising by assumptions on the category),
  and at a more general level in~\cite[Thm.~5.20.2]{wmkd20reachability}, assuming only that $\M \subseteq \Mono$.
  Here, we make no assumptions about~$\M$. Still, we follow the proof of~\cite[Thm.~5.20.2]{wmkd20reachability}, but
  the current proof differs in that
  we do not (and cannot) use that the morphisms in~$\M$ are monic, which
  was used in the final steps of the earlier proof. To address this, the
  constructed sequence of morphisms~$d_k$
  in the proof satisfies a tighter property (which does not involve $m'$).
\end{rem}
\begin{proof}[Proof of \cref{thm:reach}]
  For easier comparison, we stick close to the notation of the
  proof of~\cite[Thm.~5.20.2]{wmkd20reachability} wherever possible.
  Let $h\colon (S,s,i_S) \monoto (R, r, i_R)$ be a pointed subcoalgebra, for which we now need to show that it is a split epimorphism.
  First, we note that since $[e_k]_{k \in \N}$ is pointed coalgebra morphism, the following diagram commutes:
  \[
    \begin{tikzcd}
      I
      \arrow{r}{i_C'}
      \arrow{drr}[swap]{i_R}
      & C_0
      \arrow{r}{\inj_0}
      & \coprod_{k \in \N} C_k
      \arrow{r}[pos=0.45]{\coprod_{k \in \N} c_k}
      \arrow{d}{[e_k]_{k \in \N}}
      &[3mm] \coprod_{k \in \N} FC_{k+1}
      \arrow{r}{[F\inj_{k+1}]_{k \in \N}}
      &[6mm] F\coprod_{k \in \N} C_k
      \arrow{d}{F[e_k]_{k \in \N}}
      \\
      && R
      \arrow{rr}[swap]{r}
      && FR
    \end{tikzcd}
  \]
  Precomposing the right square with~$\inj_k$ and using properties of coproducts, this implies that the following two diagrams commute:
  \begin{equation}\label{eq:R-hom}
    \begin{tikzcd}
      I
      \arrow{r}{i_C'}
      \arrow{dr}[swap]{i_R}
      & C_0
      \arrow{d}{e_0}
      \\
      & R
    \end{tikzcd}
    \quad\&\quad
    \begin{tikzcd}
      C_k
      \arrow{r}{c_k}
      \arrow{d}[swap]{e_k}
      & FC_{k+1}
      \arrow{d}{Fe_{k+1}}
      \\
      R
      \arrow{r}[swap]{r}
      & FR
    \end{tikzcd}
  \end{equation}

  In the following, we will define a sequence of morphisms $d_k\colon C_k\rightarrowtail S$ in~$\M$ ($k \in \N$) satisfying
  \begin{equation}\label{eq:dk}
    \begin{tikzcd}
      C_k
      \arrow[>->]{r}{e_k}
      \arrow[>->]{d}[swap]{d_k}
      & R
      \\
      S
      \arrow[>->]{ur}[swap]{h}
    \end{tikzcd}
    \quad\text{and}\quad
    \begin{tikzcd}
      C_k
      \arrow{r}{c_k}
      \arrow{d}[swap]{d_k}
      & F C_{k+1}
      \arrow{d}{F d_{k+1}}
      \\
      S
      \arrow{r}[swap]{s}
      & FS
    \end{tikzcd}
    \qquad\text{for all }k\in \N.
  \end{equation}
  Note that $[e_k]_{k\in \N} \in \E$, but $e_k$ (for each $k\in \N$) is in~$\M$
  because $m_k = m' \cdot e_k$ and $m_k, m'\in \M$ (see \cref{remCancel}). Thus, every such~$d_k$ will
  necessarily be in~$\M$ as well since $h \cdot d_k = e_k$.

  Formally, we prove by induction that there exist morphisms~$d_k$ ($k \in \N$) such that for all~$k\in \N$, we have $h\cdot d_k =
  e_k$ and, if $k \ge 1$, also $s\cdot d_{k-1} = Fd_k\cdot c_{k-1}$.
  We define $d_0\colon C_0\rightarrowtail S$ using the diagonal fill-in
  property: 
  \[
    \begin{tikzcd}[sep=6mm]
      I
      \arrow[->>]{r}{i_C'}
      \arrow{d}[swap]{i_S}
      & C_0
      \arrow[>->]{d}{e_0}
      \arrow[->,dashed]{dl}[description]{d_0}
      \\
      S \arrow[>->]{r}[swap]{h}
      & R
    \end{tikzcd}
  \]
  In this diagram, the outside commutes since $i_R = e_0 \cdot i_C'$~\labelcref{eq:R-hom} and $h$ preserves the point.

  In the inductive step, given $d_k\colon C_k\rightarrowtail S$ with $h\cdot
  d_k = e_k$, the left-hand diagram below commutes.
  Using that $c_k$ is precise and $h,e_{k+1} \in \M$, we obtain the diagonal fill-in~$d_{k+1}$ (on the right) satisfying the two desired equalities.
  \[
    \begin{tikzcd}
      C_k
      \arrow[>->]{d}[swap]{d_k}
      \arrow[bend left=20]{dr}[pos=0.7]{e_k}
      \arrow{rr}{c_k}
      \descto[shift={(-2mm,-1mm)}]{dr}{\text{I.H.}}
      &
      &
      FC_{k+1}
      \arrow{dd}{Fe_{k+1}}
      \\
      S
      \arrow{d}[swap]{s}
      \arrow[->]{r}[swap]{h}
      \descto[xshift=5mm]{dr}{Hom.}
      &
      R
      \arrow{dr}{r}
      \descto{ur}{\labelcref{eq:R-hom}}
      & {}
      \\
      FS
      \arrow[->]{rr}[swap]{Fh}
      & {}
      & FR
    \end{tikzcd}
    \overset{\exists d_{k+1}}
    \Longrightarrow
    \begin{tikzcd}
      C_k
      \arrow{d}[swap]{d_k}
      \arrow{rr}{c_k}
      & & FC_{k+1}
      \arrow{ddll}{F d_{k+1}}
      \\
      S
      \arrow{d}[swap]{s}
      \\
      FS
    \end{tikzcd}
    \&
    \begin{tikzcd}
      &
      C_{k+1}
      \arrow{d}{e_{k+1}}
      \arrow{dl}[swap,pos=0.4]{d_{k+1}}
      \\
      S
      \arrow{r}[swap]{h}
      & R
    \end{tikzcd}
  \]

  Thus, \labelcref{eq:dk} and the definition of~$d_0$ yield that $[d_k]_{k\in \N}$ is a pointed coalgebra morphism, making the following diagram commute:
  \[
    \begin{tikzcd}
      I
      \arrow{r}{i_C'}
      \arrow{drr}[swap]{i_S}
      & C_0
      \arrow{r}{\inj_0}
      & \coprod_{k\in \N} C_k
      \arrow{r}[pos=0.45]{\coprod_{k\in \N}c_k}
      \arrow{d}{[d_{k}]_{k\in \N}}
      &[3mm] \coprod_{k \in \N} FC_k
      \arrow{r}{[F\inj_{k+1}]_{k\in \N}}
      &[6mm] F \coprod_{k\in \N} C_k
      \arrow{d}{F[d_{k}]_{k\in \N}}
      \\
      && S
      \arrow{rr}[swap]{s}
      & & FS
    \end{tikzcd}
  \]
  Thus, we have the following commutative square in the \emph{category of pointed coalgebras}:
  \[
    \begin{tikzcd}
      \coprod_{k\in \N}C_k
      \arrow{d}[swap]{[d_k]_{k\in \N}}
      \arrow[->>]{r}[pos=0.4]{[e_k]_{k \in \N}}
      & R
      \arrow[equals]{d}
      \arrow[dashed]{dl}[description]{g}
      \\
      S
      \arrow[>->]{r}[swap]{h}
      & R
    \end{tikzcd}
  \]
  Using the factorization system, we obtain a diagonal fill-in $g\colon R\to S$ that satisfies $h\cdot g =
  \id_{R}$, showing that $h$ is indeed a split epimorphism of coalgebras.
\end{proof}

\begin{cor}\label{corReach}
  A coalgebra~$C$ is $\M$-reachable iff the morphism~$m'$ from \cref{thm:reach} is an isomorphism.
\end{cor}
\begin{proof}
  The \textqt{if}-direction holds because $\M$-reachability is invariant under isomorphism.
  For \textqt{only if}, suppose $C$ is $\M$-reachable. 
  Then, by \itemref{reachableUnique}{isoBetweenReachable}, $m'$ is an isomorphism.
\end{proof}

If $M \subseteq \Mono$, \cref{admitPreciseIffHasLeast} shows that
precise factorizations and least bounds of morphisms coincide.
Thus, in this case, \cref{levelConstruction} instantiates exactly to
the construction of reachable subcoalgebras via least bounds introduced in~\cite[Cons.~5.19]{wmkd20reachability}.

\begin{rem}\label{exCanGraph}
  For every set functor $F\colon \Set\to\Set$, there is a canonical way
  to transform an $F$-coalgebra into a $\Pow$-coalgebra,
  which is interpreted as a directed graph and called
  the \emph{canonical graph} of the coalgebra~\cite{gumm05filter}.
  Whenever $F$ has least bounds \wrt $\M\subseteq \Mono$ and
  maps $\M$-morphisms to monos, an $F$-coalgebra is reachable
  iff its canonical graph is reachable as a
  $\Pow$-coalgebra~\cite[Thm.~4.6]{wmkd20reachability},
  \ie in the usual graph-theoretic sense.
  In such a case, the reachable subcoalgebra can simply be
  computed on the canonical graph.
\end{rem}

\bknote{}%
\jrnote{}

\section{Trees}
\label{sec:trees}
We instantiate $\M$-reachability for the non-proper
$(\Iso,\Mor)$-factorization system that exists on every category $\C$, 
resulting in an abstract characterization of tree structure in coalgebras.
To make this self-contained, we spell out the definitions and constructions explicitly.

\subsection{Universal Property}
For $\M:=\Mor$, reachability (\cref{defReachable}) instantiates to the following:
\begin{defi}\label{defTree}\hfill
  \begin{enumerate}
  \item An \emph{unravelling} of a pointed coalgebra $(C,c,i_C)$
  is a pointed coalgebra homomorphism $h\colon (T,t,i_T)\to (C,c,i_C)$.
  \item A coalgebra $(C,c,i_C)$ is called a \emph{tree} if every unravelling
    $h$ is a split epimorphism of pointed coalgebras.
  \item A \emph{tree unravelling} of a coalgebra is an unravelling that is a tree.
  \end{enumerate}
\end{defi}
\twnote{}%
\twnote{}%

\noindent Intuitively, the definition expresses that a tree (in the usual graph-theoretic
sense) is a graph that cannot be unravelled any further, because it has
no loops and no nodes share descendants.
Unsurprisingly, if a coalgebra contains a cycle, then its tree unravelling is
infinite (see \eg \cref{figFoldLoop,figSigmaUnravel}).

\begin{exa}\label{exWhySplit}
  Let us consider trees for the functor~$\FBinTree$
  from \itemref{ex:coalgebras}{ex:bintree}.
  \Cref{figTreeUnravels} shows a simple unravelling of the $\FBinTree$-coalgebra~$E$
  that sends both $q$~and~$r$ in~$D$ to the same node~$q$ in~$E$.
  There is no coalgebra morphism $E\to D$, which shows that $E$ is not a tree
  by the above definition.
  The coalgebra~$D$ \emph{is} a tree, however.
  For instance, the coalgebra~$C$ contains two copies of~$D$,
  so $C \cong D+D$ (as plain coalgebras), and the homomorphism
  $\nabla=[\id_D,\id_D]\colon D+D\to D$ has a right inverse, namely, $\inl$
  (assuming that the left-hand tree corresponds to the left-hand component;
  $\inr\colon D\to D+D$ does not preserve the point).
  Observe that $\nabla$ is not an isomorphism, which showcases why
  \cref{defReachable} of $\M$-reachability only requests
  $\M$-subcoalgebras to be split epimorphisms instead of isomorphisms.
\end{exa}

\begin{figure}
      \begin{tikzpicture}[coalgebra,baseline={(0,-0.7)}]
        \node[state,initial above] (q0) {$p$};
        \node[state] (q1) at (-0.3,-1.3) {$q$};
        \node[state] (q2) at (0.3,-1.3) {$r$};
        \draw[transition] (q0) to node[above left] {}(q1);
        \draw[transition] (q0) to node[below left] {} (q2);
        \begin{scope}[xshift=14mm]
        \node[state] (q3) {$t$};
        \node[state] (q4) at (-0.3,-1.3) {$u$};
        \node[state] (q5) at (0.3,-1.3) {$v$};
        \end{scope}
        \draw[transition] (q3) to node[above left] {}(q4);
        \draw[transition] (q3) to node[below left] {} (q5);
        \begin{scope}[on background layer]
          \node[coalgebra frame,fit={(current bounding box)}] (frame) {};
        \end{scope}
        \node[coalgname] at (frame.north west) {$C$};
      \end{tikzpicture}%
      \begin{tikzcd}
        {\!\!} \arrow[shift left=2]{r}{\nabla}
        & {}
        {\!\!} \arrow[shift left=2]{l}{\inl}
      \end{tikzcd}%
      \begin{tikzpicture}[coalgebra,baseline={(0,-0.7)}]
        \node[state, initial above] (q0) {$p$};
        \node[state] (q1) at (-0.3,-1.3) {$q$};
        \node[state] (q2) at (0.3,-1.3) {$r$};
        \draw[transition] (q0) to node[above left] {}(q1);
        \draw[transition] (q0) to node[below left] {} (q2);
        \begin{scope}[on background layer]
          \node[coalgebra frame,fit={(current bounding box)}] (frame) {};
        \end{scope}
        \node[coalgname] at (frame.north west) {$D$};
        \begin{scope}[commutative diagrams/.cd,every diagram]
          \path[commutative diagrams/every arrow,
            shorten > = 1mm,
            shorten < = 1mm]
            (frame)
            to[loop,out=80,in=100,looseness=4]
              node[commutative diagrams/every label,above]
                {\ensuremath{s}}
            (frame);
        \end{scope}
      \end{tikzpicture}%
      \begin{tikzcd}
        {\!\!} \arrow[shift left=2]{r}{t} &
        {\!\!} \arrow[shift left=2]{l}[anchor=center]{/}
      \end{tikzcd}%
      \begin{tikzpicture}[coalgebra,baseline={(0,-0.7)}]
        \node[state,initial above] (q0) {$p$};
        \node[state] (q1) at (0,-1.3) {$q$};
        \draw[transition,bend left] (q0) to node[above left] {}(q1);
        \draw[transition,bend right] (q0) to node[below left] {} (q1);
        \begin{scope}[on background layer]
          \node[coalgebra frame,fit={(current bounding box)}] (frame) {};
        \end{scope}
        \node[coalgname] at (frame.north west) {$E$};
      \end{tikzpicture}%
    \caption{Unravelling for $\FBinTree X=X\times X + \set{\bot}$.}
    \label{figTreeUnravels}
\end{figure}

\begin{rem}
  Note that $E$ from \cref{figTreeUnravels} shows that even if the canonical
  graph of a coalgebra is a tree (in the graph-theoretic sense), the coalgebra itself might not be a tree
  (in the sense of \cref{defTree}).
  The canonical graph of $E$ is simply $\to p\longrightarrow q$, which is a tree.
\end{rem}

In the example above, $\inl\colon D\to C$ is not surjective and thus not a split epimorphism of
coalgebras, which shows that $C$ is not a tree. This is no coincidence, because
trees are always reachable:
\begin{lem}\label{lemTreeReachable}
  For any class $\M$ of morphisms, every tree is $\M$-reachable.
\end{lem}
\begin{proof}
  Let $(C,c,i_C)$ be a tree, and $h \colon (T,t,i_T) \monoto (C,c,i_C)$ a pointed $\M$-subcoalgebra. Then $h$ is immediately a
  split epimorphism of pointed coalgebras as $(C,c,i_C)$ is a tree.
\end{proof}
There is a slight variation of the definition of trees in terms of reachable coalgebras:
\begin{defi}\label{defUnfolded}
  For a class of morphisms $\M$,
  a pointed coalgebra $C\in \Coalg_I(F)$ is called \emph{($\M$-)reachably unfolded}
  if every pointed coalgebra morphism $h\colon R\to C$ from an $\M$-reachable coalgebra
  $R\in\Coalg_I(F)$ is an isomorphism.
\end{defi}

Under some conditions, trees coincide with $\M$-reachably unfolded coalgebras:
\begin{thm}\label{thmReachablyUnfolded}\hfill
  \begin{enumerate}
  \item If $\SplitMono \subseteq \M$, then every tree is $\M$-reachably unfolded.
  \item\label{thmReachablyUnfoldedToTree} If every pointed coalgebra admits a morphism from some $\M$-reachable coalgebra, then every
  $\M$-reachably unfolded coalgebra is a tree (and thus also
  $\M$-reachable).
  \end{enumerate}
\end{thm}
\noindent The assumptions of \itemref{thmReachablyUnfolded}{thmReachablyUnfoldedToTree} are satisfied in particular when every pointed coalgebra has an $\M$-reachable subcoalgebra, as in the setting of \cref{sec:reachablePartCons}.\bknote{}%
\begin{proof}\hfill
\begin{enumerate}
\item
    Let $(C,c,i_C)$ be a tree, and $h \colon (R,r,i_R) \to (C,c,i_C)$ a
    homomorphism from an $\M$-reachable $(R,r,i_R)$. As $C$ is a tree, $h$ is a
    split epimorphism so that it has a section $s \colon (C,c,i_C) \to (R,r,i_R)$.
    Since sections are split mono by definition and $s$ is a pointed
    coalgebra homomorphism, $s$ is a pointed $\M$-subcoalgebra of
    $R$. Thus, $s$ is also split epi, and so an isomorphism. Then, also
    $h$ must be an isomorphism. Hence, $C$ is reachably unfolded.

\item
    Let $(C,c,i_C)$ be reachably unfolded and $h \colon (T,t,i_T) \to (C,c,i_C)$
    some unravelling. Let $h' \colon (R,r,i_R) \to (T,t,i_T)$ be a morphism
    from an $\M$-reachable coalgebra $R$. By assumption, $h \cdot h'$
    must be an isomorphism, so that it has an inverse, say $s \colon C \to R$.
    Then, $h' \cdot s$ is a section of $h$ as
    $h \cdot (h' \cdot s) = (h \cdot h') \cdot s = \id_C$. Further, $h'$ and $s$
    are pointed coalgebra homomorphisms, so their composition is as well.

    For the additional comment on $\M$-reachability,
    note that since $(C,c,i_C)$ is a tree, it is also $\M$-reachable by \cref{lemTreeReachable}.
    \qedhere
\end{enumerate}
\end{proof}

\begin{rem}
The notion of $\M$-reachable unfoldedness degrades if
the assumptions of \cref{thmReachablyUnfolded} are not met, and then $\M$-reachable unfoldedness differs from being a tree:
\begin{enumerate}
\item
Let $\M = \Iso$ (note that $\SplitMono \nsubseteq \M$). Then every coalgebra is
$\M$-reachable and thus a pointed coalgebra $D$ is reachably unfolded iff every coalgebra morphism $C\to D$ is an isomorphism.
Thus, no coalgebra in $\Set$ is reachably unfolded for $\M = \Iso$, because
$\nabla = [\id_D,\id_D] \colon D + D\to D$ is not an isomorphism.
(The empty set carries no pointed coalgebra structure.)
\item
Let $\M = \Mor$, \ie the class of all morphisms in $\C$
(note that \cref{defReachable,defUnfolded} impose no assumptions on $\M$).
Then $\M$-reachability is the property of being a tree. There are functors $F$
for which there is an $F$-coalgebra $(C,c,i_C)$ without any tree unravelling (for instance, $F = \Pow$;
see \cref{exNoPowTree}). Such a $(C,c,i_C)$ is in
particular not a tree, but it is $\M$-reachably unfolded, for the trivial reason that
no morphism from a tree (\ie $\M$-reachable coalgebra) exists.
\end{enumerate}
\end{rem}

\noindent From the generalized uniqueness result (\itemref{reachableUnique}{isoBetweenReachable}), we obtain that tree unravellings are unique up to isomorphism:
\begin{lem}\label{treeUnique}\hfill
  \begin{enumerate}
  \item\label{treeIso} Every coalgebra morphism $h\colon C\to D$ between trees is an isomorphism.
  \item If $\C$ has pullbacks and $F$ preserves them weakly, then every
    coalgebra has at most one tree unravelling.
  \end{enumerate}
\end{lem}
\begin{proof}
  Instantiate \cref{reachableUnique} to the $(\Iso,\Mor)$-factorization system. Then
  $\liftfactsys{\M}{F}$-morphisms are arbitrary morphisms and $\M$-intersections are
  pullbacks.
\end{proof}

\begin{rem}\label{exWhyNotUniqu}
In contrast to other categorical uniqueness results (\eg that of (co)limits),
tree unravellings are not unique up to \emph{unique} isomorphism, because
isomorphic but distinct subtrees can always be permuted by automorphisms. For
example, consider the automorphism $s\colon D\to D$ in \cref{figTreeUnravels} that
swaps $q$ and $r$; since $s\neq \id_D$, the tree unravelling $D$ of $E$ is not
unique up to unique isomorphism.
\end{rem}

Let us look at a few more examples of trees. We start with partial deterministic automata.

\begin{defi}\label{defInputCoalg}
  For a partial deterministic automaton $\fpair{o,\delta}\colon C\to O\times (C+\set{\bot})^A$ with a point $q_0\colon 1\to C$,
  define the \emph{coalgebra of defined inputs} $\fpair{\bar o, \bar \delta}$ by
  \[
    P = \set{w\in A^*\mid \exists q\in C\colon \delta^*(q_0)(w) = \inl(q)},
    \quad
    \bar o\colon P\to O
    \quad
    \bar o(w) = o(\delta^*(q_0)(w)),
    \quad
    \varepsilon\colon 1\to P,
  \]
  \[
    \bar\delta\colon P\to (P + \set{\bot})^A,
    \qquad
    \bar\delta(w)(a) = \begin{cases}
      \inl(w\,a)&\text{if }\delta(\delta^*(q_0)(w)) = \inl(q)\text{ for some $q \in C$}, \\
      \inr(\bot)&\text{otherwise.}
    \end{cases}
  \]
\end{defi}
In other words, $P$ is the domain of the partial map $\delta^*(q_0)\colon A^*\partialto C$. Thus,
$\delta^*(q_0)$ restricts to a total map $\delta^*(q_0)\colon P\to C$, which
is a pointed coalgebra morphism by construction.
\begin{lem}\label{dfaInputsIsTree}
  For every partial deterministic automaton, its coalgebra of defined inputs (\cref{defInputCoalg}) is a tree.
\end{lem}
\begin{proof}
  Since $\FAut{A}$ is built up from constants, variables, products, coproducts, and the right adjoint $(-)^A$,
  \cref{morEndoPrecise,rightAdjointPreciseFact} imply that
  it admits precise factorizations \wrt $\Mor$, and hence also \wrt $\Mono$
  by \cref{preciseFactMorM}.
  As $\FAut{A}$ also preserves monos, we can apply \cref{thm:reach}
  to conclude that every partial deterministic automaton
  has a $\Mono$-reachable subcoalgebra.
  Thus, by \itemref{thmReachablyUnfolded}{thmReachablyUnfoldedToTree},
  it suffices to show that $(P, \fpair{\bar o, \bar\delta}, \varepsilon)$ is $\Mono$-reachably unfolded.

  We need to show that
  every coalgebra morphism $h\colon (R,\fpair{o_R,\delta_R},r_0)\longrightarrow (P,\fpair{\bar o,\bar\delta},\varepsilon)$ from a reachable coalgebra $R$ to the coalgebra of
  defined inputs of $(C,\fpair{o,\delta},q_0)$ is bijective.
  Given that $P$ is reachable (\cf \cref{dfaReachable}) and $\FAut{A}$ preserves monos, $h$ is surjective by \cref{reachEveryE}.
  For the remaining injectivity proof,
  take two states $r_1,r_2\in R$ with $h(r_1) = h(r_2)$.
  Since $R$ is reachable, by \cref{dfaReachable}, there are $w_1,w_2\in A^*$ with
  \[
    \delta_R^*(r_0)(w_1) = \inl(r_1)
    \text{ and }
    \delta_R^*(r_0)(w_2) = \inl(r_2).
  \]
  Since $h$ is a homomorphism, this implies
  \[
    \bar\delta^*(\varepsilon)(w_1) = \inl(h(r_1))
    \text{ and }
    \bar\delta^*(\varepsilon)(w_2) = \inl(h(r_2)).
  \]
  Whenever $\bar\delta^*(\varepsilon)(v) = \inl(u)$, then $u = v$ by \cref{defInputCoalg}. Hence, $w_1 = h(r_1) = h(r_2) = w_2$, which implies $r_1 = r_2$.
\end{proof}
\begin{cor}\label{corDFAUnravelling}
  The tree unravelling of a partial deterministic automaton is given by its coalgebra of defined inputs (\cref{defInputCoalg}).
\end{cor}
\begin{proof}
  Immediate from \cref{dfaInputsIsTree}.
\end{proof}

\begin{thm}\label{thmDFATree}
  For every partial deterministic automaton~$(C, \fpair{o, \delta}, q_0)$, the following are equivalent:
  \begin{enumerate}
  \item\label{thmDFATreeCoalg} $(C, \fpair{o, \delta} , q_0)$ is a tree (in the coalgebraic sense).
  \item\label{thmDFATreeIso} The morphism $\delta^*(q_0) \colon P \to C$ from the coalgebra of defined inputs is an isomorphism.
  \item\label{thmDFATreeUniquePath} For every~$q \in C$, there is precisely one $w \in A^*$ such that $\delta^*(q_0)(w) = \inl(q)$.
  \end{enumerate}
\end{thm}
\begin{proof}\hfill
  \begin{itemize}
  \item \labelcref*{thmDFATreeCoalg} $\Rightarrow$ \labelcref*{thmDFATreeIso}:
    Since $P$~and~$C$ are both trees, $\delta^*(q_0)$ is an isomorphism by \itemref{treeUnique}{treeIso}.
  \item \labelcref*{thmDFATreeIso} $\Rightarrow$ \labelcref*{thmDFATreeCoalg}:
    $C$ is isomorphic to~$P$, which is a tree by \cref{dfaInputsIsTree}.
    Thus, $C$ is also a tree.
  \item \labelcref*{thmDFATreeIso} $\Leftrightarrow$ \labelcref*{thmDFATreeUniquePath}:
    Follows immediately from the characterization of bijections as having unique preimages.
  \qedhere
  \end{itemize}
\end{proof}

\noindent We now take a look at graphs.
\begin{defi}\label{defPathCoalg}
  For a pointed multigraph $G = (V,E,s,t,v_0)$,
  define the \emph{coalgebra of rooted paths} by
  \[
    \bar E = \bigcup_{v\in V} \Path(v_0,v),
    \qquad
    d\colon \bar{E}\to \Bag \bar{E},
    \qquad
    d(p)(p') = \begin{cases}
      1 &\text{if there is }e\in E\text{ with }(p,e) = p', \\
      0 &\text{otherwise.}
    \end{cases}
  \]
  The distinguished point $i_D = \varepsilon$ is the path of length 0 (from $v_0$ to $v_0$).
\end{defi}

\begin{prop}
\label{pathCoalgMorphism}
  For each pointed multigraph $G$,
  the map $t\colon \bar E \to V$ sending each path to its target vertex
  is a pointed coalgebra morphism from the coalgebra of rooted paths to $G$
  considered as a $\Bag$-coalgebra.
\end{prop}
\begin{proof}
  The map $t$ sends the empty path on $v_0$ to $v_0$. For the verification that the square
  \[
    \begin{tikzcd}
      \bar E
      \arrow{r}{d}
      \arrow{d}[swap]{t}
      & \Bag \bar E
      \arrow{d}{\Bag t}
      \\
      V
      \arrow{r}[swap]{c}
      & \Bag V
    \end{tikzcd}
  \]
  commutes, take $p\in \bar{E}$ and $u\in V$:
  \begin{align*}
  \allowdisplaybreaks
    (\Bag t\cdot d)(p)(u)
    &= \Bag t(d(p))(u)
    = \sum_{\substack{p' \in \bar{E}\\ t(p') = u}} d(p)(p')
    \\
    &= \sum_{\substack{p' \in \bar{E}\\ t(p') = u}} \begin{cases}
      1 & \text{if there is some }e\in E\text{ with } (p,e)=p'\\
      0 &\text{otherwise}
    \end{cases}
    \\
    &= |\set{p'\in \bar{E}\mid t(p') = u, \exists e\in E\colon (p,e)=p'}|
    \\
    &= |\set{e\in E\mid t(p) = s(e), t(e) = u}|
    \\
    &= c(t(p))(u)
    = (c\cdot t)(p)(u).
  \qedhere
  \end{align*}
\end{proof}

\begin{lem}\label{lemPathsTree}
  For every pointed multigraph, its coalgebra of rooted paths (\cref{defPathCoalg}) is a tree (in the coalgebraic sense).
\end{lem}
\begin{proof}
  Since $\Bag$ admits precise factorizations \wrt $\Mor$ by \cref{bagPreciseFact},
  it does so \wrt $\Mono$ as well by \cref{preciseFactMorM}.
  As $\Bag$ also preserves monos, we can apply \cref{thm:reach}
  to conclude that every pointed multipgraph
  has a $\Mono$-reachable subcoalgebra.
  Thus, by \itemref{thmReachablyUnfolded}{thmReachablyUnfoldedToTree},
  it suffices to show that $(\bar{E}, d, \varepsilon)$ is $\Mono$-reachably unfolded.
  
  Given a homomorphism from a reachable $\Bag$-coalgebra $h\colon (R,r,i_R)\to
  (\bar E, d,\varepsilon)$, we show that $h$ is bijective.
  Denote the multigraph associated to $R$ by $E_R$, with maps $s_R,t_R\colon E_R\to R$ and root $q_0$.
  Surjectivity is direct from the reachability of $\bar{E}$ (\cf \cref{bagReachable}) using the fact that $\Bag$ preserves monos and \cref{reachEveryE}.
  For injectivity, consider $x_1,x_2\in R$ with $h(x_1) = h(x_2)$. Given that
  $R$ is reachable, by \cref{bagReachable}, there are paths $p_1,p_2$ in $R$ with $t_R(p_1) = x_1$ and
  $t_R(p_2) = x_2$.
  We prove $x_1 = x_2$ by induction over the length of the paths $p_1,p_2$, also
  universally quantifying over $x_1,x_2$. We prove that
  \[
    \text{for all $x_1,x_2\in R$, $p_1\in \Path(q_0,x_1)$, $p_2\in \Path(q_0,x_2)$, if $h(x_1) = h(x_2)$, then $x_1=x_2$.}
  \]
  \begin{itemize}
  \item If both $p_1$ and $p_2$ have length 0, then we directly have $x_1=q_0=x_2$, as desired.
  \item If one of the paths has length 0 and the other is nonempty, let $p_1$ be empty (w.l.o.g), implying
    $q_0 = x_1$. Thus, $p_2$ is a nonempty path from $x_1$ to $x_2$. The
    homomorphism turns this path into a non-empty path from $h(x_1)$ to $h(x_2)
    = h(x_1)$ in the coalgebra of rooted paths, which contradicts \cref{defPathCoalg}.
  \item Consider paths $p_1=(p_1',e_1) \in \Path(q_0,x_1)$ and  $p_2=(p_2',e_2) \in \Path(q_0,x_2)$.
    From $h(t_R(e_1)) = h(t_R(p_1)) = h(x_1) = h(x_2) = h(t_R(p_2)) = h(t_R(e_2))$, \cref{defPathCoalg}
    implies that $h(s_R(e_1)) = h(s_R(e_2))$.
    By the induction hypothesis, we obtain $s_R(e_1) = s_R(e_2)$.
    Thus, $x_1$ and $x_2$ are necessarily equal. Because if they were not,
    then the $\Bag$-coalgebra homomorphism property enforces that the
    multiplicities of the edges $e_1$ and $e_2$ from $y:=s_R(e_1) = s_R(e_2)$ to $x_1$
    and $x_2$, respectively, are added up to a multiplicity of at least 2 for the edge from $h(y)$
    to $h(x_1)=h(x_2)$, which contradicts \cref{defPathCoalg}.
    \qedhere
  \end{itemize}
\end{proof}

\begin{cor}\label{corBagUnravelling}
  The tree unravelling of a multigraph is given by its coalgebra of rooted paths (\cref{defPathCoalg}).
\end{cor}
\begin{proof}
  Immediate from \cref{pathCoalgMorphism,lemPathsTree}.
\end{proof}

\begin{thm}\label{thmBagTree}
  For every $\Bag$-coalgebra~$(V,c,i_V)$, the following are equivalent:
  \begin{enumerate}
  \item\label{thmBagTreeCoalg} $(V,c,i_V)$ is a tree (in the coalgebraic sense).
  \item\label{thmBagTreeIso} The morphism $t\colon \bar E\to V$ from the coalgebra of rooted paths is an isomorphism.
  \item\label{thmBagTreeUniquePath} For every~$v\in V$, there is precisely one path from~$v_0$ to~$v$.
  \end{enumerate}
\end{thm}
\begin{proof}\hfill
  \begin{itemize}
  \item \labelcref*{thmBagTreeCoalg} $\Rightarrow$ \labelcref*{thmBagTreeIso}:
    Since $\bar{E}$~and~$V$ are both trees, $t$ is an isomorphism by \itemref{treeUnique}{treeIso}.
  \item \labelcref*{thmBagTreeIso} $\Rightarrow$ \labelcref*{thmBagTreeCoalg}:
    $V$ is isomorphic to~$\bar E$, which is a tree by \cref{lemPathsTree}.
    Thus, $V$ is also a tree.
  \item \labelcref*{thmBagTreeIso} $\Leftrightarrow$ \labelcref*{thmBagTreeUniquePath}:
    Follows immediately from the definition of $t\colon \bar E \to V$ and the characterization of bijections as having unique preimages.
  \qedhere
  \end{itemize}
\end{proof}
\noindent The equivalence \labelcref*{thmBagTreeCoalg} $\Leftrightarrow$ \labelcref*{thmBagTreeUniquePath} in \cref{thmBagTree} shows in particular
that a $\Bag$-coalgebra is a tree in the coalgebraic sense
iff it is a tree in the usual sense of a directed graph. 

\begin{exa}
  In \cref{unravelBag}, the coalgebra morphism $h\colon C\to D$ is the tree
  unravelling of the $\Bag$-coalgebra~$D$. Note that for every vertex~$x\in D$,
  the number of paths from the root~$r$ to~$x$ determines the number of copies
  of~$x$ in the tree unravelling~$C$. For instance, there are three paths to~$q$: one
  direct edge from~$r$ and two paths via~$p$.
\end{exa}
\begin{figure}[h]
  \begin{tikzpicture}[x=7mm,y=8mm]
    \begin{scope}[coalgebra,scope of math nodes,local bounding box=cyclic]
      \node[state,initial above] (r) {r};
      \node[state] (p) at (-.8,-1.2) {p};
      \node[state] (q) at (1,-2.0) {q};
      \node[state] (v) at (-.8,-3.2) {v};
      \path[transition] (r) to (p);
      \path[transition] (r) to (q);
      \path[transition] ([yshift=0mm]p.east) to ([yshift=2mm]q.west);
      \path[transition] ([yshift=-2mm]p.east) to ([yshift=0mm]q.west);
      \path[transition] (p) to (v);
      \path[transition] (q) to (v);
    \end{scope}
    \begin{scope}[xshift=-52mm,coalgebra,scope of math nodes,
        sibling distance=7mm,level distance=9mm,
        y=7mm,
        state/.append style={inner sep=0pt},
        local bounding box=tree]
      \node[state,initial above] (r1) {r};
      \node[state] (p1) at (-1,-1) {p}
        child {node[state] (v1) {v_1} edge from parent[transition] }
        child {node[state] (q2) {q_2} edge from parent[transition]
          child {node[state] (v2) {v_2} edge from parent[transition] }
          }
        child {node[state] (q3) {q_3} edge from parent[transition]
          child {node[state] (v3) {v_3} edge from parent[transition] }
          }
        ;
      \node[state] (q1) at (1,-1) {q_1}
        child {node[state] (v4) {v_4} edge from parent[transition] }
        ;
      \path[transition] (r1) to (p1);
      \path[transition] (r1) to (q1);
    \end{scope}
    \begin{scope}[on background layer]
      \node[coalgebra frame,fit=(cyclic)] (cyclic frame) {};
      \node[coalgname] at (cyclic frame.north west) {\ensuremath{D}};
      \node[coalgebra frame,fit=(tree)] (tree frame) {};
      \node[coalgname] at (tree frame.north west) {\ensuremath{C}};
    \end{scope}
      \begin{scope}[commutative diagrams/.cd,every diagram]
        \path[commutative diagrams/every arrow,
          shorten > = 2mm,
          shorten < = 2mm]
          (tree frame.east)
          to node[commutative diagrams/every label] {\ensuremath{h}}
          node[commutative diagrams/every label,below]
            {\ensuremath{\substack{
              r\mapsto r\\
              p\mapsto p\\
              q_i\mapsto q\\
              v_i\mapsto v
              }}}
          (cyclic frame.west |- tree frame.east);
      \end{scope}
  \end{tikzpicture}
    \caption{Unravelling of a $\Bag$-coalgebra.}
    \label{unravelBag}
\end{figure}

One might first be tempted to model graphs as $\Pow$-coalgebras. However,
$\Pow$-coalgebras do not provide the right notion of trees:
\begin{exa}\label{exNoPowTree}
  If a $\Pow$-coalgebra is a tree, then it has one vertex and no edges.
  In other words, no $\Pow$-coalgebra with an edge is a tree.
  The intuitive reason is that the multiplicity of elements in a set
  does not matter, so it is always possible to duplicate siblings.
  For a concrete example, see \cref{fig:noPowTree}.
  Since $\set{q} = \set{q,q}$, the coalgebras $D$~and~$E$ are identical (in a
  pure set-theoretic sense). The coalgebra~$D$ is not a tree, as witnessed by
  the unravelling to~$C$. Now $C$ is still not a tree, because we can
  apply the same principle again to duplicate the transition from~$p$ to~$q$.
  As a consequence, no $\Pow$-coalgebra with an edge has a tree unravelling.
  \begin{figure}[h]
      \begin{tikzpicture}[coalgebra,baseline={(0,-0.7)}]
        \node[state,initial above] (q0) {$p$};
        \node[state] (q1) at (-0.5,-1.3) {$q$};
        \node[state] (q2) at (0.5,-1.3) {$r$};
        \draw[transition] (q0) to node[above left] {}(q1);
        \draw[transition] (q0) to node[below left] {} (q2);
        \begin{scope}[on background layer]
          \node[coalgebra frame,fit={(current bounding box)}] (last frame) {};
        \end{scope}
        \node[coalgname] at (last frame.north west) {$C$};
      \end{tikzpicture}%
      \begin{tikzcd}[column sep=15mm]
        {} \arrow[->>]{r}[above]{\displaystyle h}[below,yshift=-2pt]{%
          \substack{%
          p \mapsto p \\
          q \mapsto q \\
          r \mapsto q \\
          }}
        & {}
      \end{tikzcd}%
      \begin{tikzpicture}[coalgebra,baseline={(0,-0.7)}]
        \node[state,initial above] (q0) {$p$};
        \node[state] (q1) at (0,-1.3) {$q$};
        \draw[transition,bend left=20] (q0) to node[above] {}(q1);
        \draw[transition,bend right=20] (q0) to node[below] {} (q1);
        \begin{scope}[on background layer]
          \node[coalgebra frame,inner xsep=4mm,fit={(current bounding box)}] (last frame) {};
        \end{scope}
        \node[coalgname] at (last frame.north west) {$D$};
      \end{tikzpicture}%
      ~$=$~%
      \begin{tikzpicture}[coalgebra,baseline={(0,-0.7)}]
        \node[state,initial above] (q0) {$p$};
        \node[state] (q1) at (0,-1.3) {$q$};
        \draw[transition,bend left=0] (q0) to node[above] {}(q1);
        \begin{scope}[on background layer]
          \node[coalgebra frame,inner xsep=5mm,fit={(current bounding box)}] (last frame) {};
        \end{scope}
        \node[coalgname] at (last frame.north west) {$E$};
      \end{tikzpicture}
    \caption{Unravelling a $\Pow$-coalgebra.}
    \label{fig:noPowTree}
  \end{figure}
\end{exa}
\begin{exa}
  In the case of~$\Bag$, the above issue of transition duplication
  (\cref{exNoPowTree}) cannot happen, as visualized in \cref{fig:unravelFlatBag}.
  In the coalgebra~$E$, the transition from~$p$ to~$q$ has multiplicity~$2$,
  which we understand as $E$ being a multigraph with two edges (visualized as~$D$).
  Thus, $q$ can be duplicated at most once, resulting in the coalgebra~$C$.
  Every coalgebra morphism from a reachable coalgebra to~$C$ is then
  necessarily an isomorphism.
  \begin{figure}[h]
    \begin{tikzpicture}[coalgebra,baseline={(0,-0.7)}]
      \node[state,initial above] (q0) {$p$};
      \node[state] (q1) at (-0.5,-1.3) {$q$};
      \node[state] (q2) at (0.5,-1.3) {$r$};
      \draw[transition] (q0) to node[on transition] {1} (q1);
      \draw[transition] (q0) to node[on transition] {1} (q2);
      \begin{scope}[on background layer]
        \node[coalgebra frame,fit={(current bounding box)}] (last frame) {};
      \end{scope}
      \node[coalgname] at (last frame.north west) {$C$};
    \end{tikzpicture}%
    \begin{tikzcd}[column sep=15mm]
      {} \arrow[->>]{r}[above]{\displaystyle h}[below,yshift=-2pt]{%
        \substack{%
        p \mapsto p \\
        q \mapsto q \\
        r \mapsto q \\
        }}
      & {}
    \end{tikzcd}%
    \begin{tikzpicture}[coalgebra,baseline={(0,-0.7)}]
      \node[state,initial above] (q0) {$p$};
      \node[state] (q1) at (0,-1.3) {$q$};
      \draw[transition,bend left=20] (q0) to node[above] {}(q1);
      \draw[transition,bend right=20] (q0) to node[below] {} (q1);
      \begin{scope}[on background layer]
        \node[coalgebra frame,inner xsep=4mm,fit={(current bounding box)}] (last frame) {};
      \end{scope}
      \node[coalgname] at (last frame.north west) {$D$};
    \end{tikzpicture}%
    ~$=$~%
    \begin{tikzpicture}[coalgebra,baseline={(0,-0.7)}]
      \node[state,initial above] (q0) {$p$};
      \node[state] (q1) at (0,-1.3) {$q$};
      \draw[transition,bend left=0] (q0) to node[on transition] {2} (q1);
      \begin{scope}[on background layer]
        \node[coalgebra frame,inner xsep=5mm,fit={(current bounding box)}] (last frame) {};
      \end{scope}
      \node[coalgname] at (last frame.north west) {$E$};
    \end{tikzpicture}
    \caption{Unravelling a $\Bag$-coalgebra.}
    \label{fig:unravelFlatBag}
  \end{figure}
\end{exa}

\subsection{Construction}
The generalized reachability construction (\cref{levelConstruction}) instantiates nicely to trees for $\E=\Iso$ and $\M = \Mor$.

\begin{asm}
  For the remainder of Section~\thesubsection, we
  fix a functor $F\colon \C\to \C$ and assume that $F$ admits precise factorizations \wrt $\Mor$.
\end{asm}
\begin{cons}\label{constrTree}
  For a pointed coalgebra $(C,c,i_C)$, we define the following sequence of maps
  $h_k\colon T_k\to C$ and precise maps $t_k\colon T_k\to F T_{k+1}$ ($k\in \N$) inductively:
    \begin{enumerate}
    \item $T_0 = I$, $h_0= i_C\colon T_0\to C$.
    \item $t_{k}$ and $h_{k+1}$ are the precise factorization of $c\cdot h_k$:
    \[
      \begin{tikzcd}
        T_k
        \arrow{d}[swap]{h_k}
        \arrow[dashed]{r}{t_k}
        & FT_{k+1}
        \arrow[dashed]{d}{Fh_{k+1}}
        \\
        C
        \arrow{r}[swap]{c}
        & FC
      \end{tikzcd}
    \]
    \end{enumerate}
    We call $T_k$ the $k$-th level
    and $[h_k]_{k\in \N}\colon \coprod_{k} T_k \longrightarrow C$ the
    \emph{coproduct of levels} of $(C,c,i_C)$.
\end{cons}
\begin{rem}
  Since $\E = \Iso$, the factorization in \autoref{levelConstruction} simplifies
  so that $C_0$ becomes $I$ and $i_C' = \id_I$.
\end{rem}

\begin{thm}\label{thm:treeUnravCon}
  For every coalgebra, the coproduct of its levels (\cref{constrTree}) is a tree unravelling.
\end{thm}
\begin{proof}
  Special case of \cref{thm:reach} for the $(\Iso, \Mor)$-factorization system.
  Note that in this factorization system,
  $R$ and $m'$ can be taken to be $\coprod_{k \in \N} C_k$ and $[m_k]_{k \in \N}$, respectively.
\end{proof}
\begin{cor}\label{corTreeCoprodLevels}
  A coalgebra is a tree iff it is isomorphic to the coproduct of its levels.
\end{cor}
\begin{proof}
  Follows immediately from the isomorphism invariance of being a tree, \cref{thm:treeUnravCon}, and \cref{treeUnique}.
\end{proof}

\ifdraft{\par\noindent
\bknote{}%
\textbf{The following variant of \cref{exCanGraph} does not work:}
\textit{
  Whenever a functor $F$ has $F$-precise factorizations,
  then an $F$-coalgebra is a tree iff its canonical graph is a tree (in the
  usual sense of directed graphs).
}
\twnote{}%
}{}

\begin{exa}\label{exFinalCoalg}
  Adámek and Porst~\cite{AdamekP04} consider the elements of the \emph{final coalgebra}
  as \emph{trees}. However, no coalgebraic definition of being a tree for a
  general functor is provided. Instead, they define a \emph{tree coalgebra}~$A_t$ for a
  polynomial functor $F_\Sigma\colon\Set\to\Set$ and an element~$t$ of the final
  $F_\Sigma$-coalgebra~$(T_\Sigma,\tau)$~\cite[II.3]{AdamekP04}.
  Such an element~$t\in T_\Sigma$ is a partial function $t\colon
  \N^*\partialto \Sigma$ that, intuitively speaking, returns the symbol~$\sigma\in \Sigma$
  whose location in the tree is described by a position~$p\in \N^*$. For~$t\in T_\Sigma$, the tree coalgebra~$A_t$ then consists of the
  domain of the partial function~$t$, which is intuitively the set of rooted
  paths in the tree, analogous to the coalgebras of rooted paths and defined
  inputs considered above.
  In particular, the coalgebra~$A_t$ is not isomorphic to the
  subcoalgebra of~$T_\Sigma$ generated by~$t$.
  We can phrase their observation using pointed coalgebras:
\end{exa}
\begin{obsC}[{\cite[II.4]{AdamekP04}}]
  For every pointed coalgebra $1\xrightarrow{x} C\xrightarrow{c} F_\Sigma C$
  there is a unique tree $t\colon \N^*\partialto \Sigma$ and pointed coalgebra
  morphism:
  \[
    \begin{tikzcd}
      1
      \arrow{r}{t}
      \arrow{rd}[swap]{x}
      & A_t
      \arrow{d}{h}
      \arrow{r}{a_t}
      & F_\Sigma A_t
      \arrow{d}{F_\Sigma h}
      \\
      & C
      \arrow{r}{c}
      & F_\Sigma C
    \end{tikzcd}
  \]
  If $(C,c)$ happens to be the final coalgebra, then $t = x$.
\end{obsC}
In the proof of \cite[II.6]{AdamekP04}, Adámek and Porst show that for every
coalgebra morphism $g\colon (B,b)\to (C,c)$ with $s\in B$, the homomorphism
$h\colon (A_t,a_t)\colon (C,c)$ factors through $g$. Considering $(B,b)$ with the point $s\colon 1\to B$
yields the projectivity criterion (\cref{thm:projective}):
\begin{cor}
  For all pointed coalgebra morphisms $g$ and $h$, there is (an even unique)
  pointed homorphism $f$:
  \[
    \begin{tikzcd}
      &
      (B,b,s)
      \arrow{d}{g}
      \\
      (A_t,a_t,t)
      \arrow{r}{h}
      \arrow[dashed]{ur}{f}
      &
      (C,c,s)
    \end{tikzcd}
  \]
\end{cor}
  So $A_t$ is indeed also \emph{the} tree unravelling of the pointed
  $F_\Sigma$-coalgebra~$(T_\Sigma, \tau, t)$.\bknote{}
\begin{exa}
  An example of an $F_\Sigma$-coalgebra is visualized in \cref{figSigmaUnravel}.
  Its tree unravelling~$T$ with levels~$(T_k)_{k\in \N}$ arising from \cref{constrTree} is shown on the left.
  The resulting coalgebra is the same as the tree coalgebra~$A_t$~\cite[II.3]{AdamekP04} where $t \colon \N^* \partialto \Sigma$ is given by
  \[
    t(p) =
      \begin{cases}
        *    & \text{if $p = 0^n$ for some~$n \in \N$,} \\
        a    & \text{if $p = 0^{2n}1$ for some~$n \in \N$,} \\
        b    & \text{if $p = 0^{2n+1}1$ for some~$n \in \N$,} \\
        \bot & \text{otherwise.}
      \end{cases}
  \]
  Note that there is a unique pointed coalgebra morphism $h\colon T\to C$.
  \begin{figure}
    \begin{tikzpicture}[coalgebra frame/.append style={
      inner sep=3pt,
    }]
    \begin{scope}[coalgebra,local bounding box=scope1,scope of math nodes,x=10mm,y=10mm]
        \node[state,initial left] (i) {*};
        \node[state] (a) at (0,-1) {a};
        \node[state] (j) at (1,0) {*};
        \node[state] (b) at (1,-1) {b};
        \path[transition] (i) to (a) ;
        \path[transition,bend right=25] (i) to (j) ;
        \path[transition] (j) to (b) ;
        \path[transition,bend right=25] (j) to (i) ;
      \end{scope}
      \begin{scope}[coalgebra,local bounding box=scope2,xshift=-8cm,scope of math nodes,x=10mm,y=10mm]
        \node[state,initial left] (i1) {*};
        \node[state] (a1) at (1,-1) {a};
        \node[state] (j1) at (1,0) {*};
        \node[state] (b1) at (2,-1) {b};
        \begin{scope}[xshift=20mm]
          \node[state] (i2) {*};
          \node[state] (a2) at (1,-1) {a};
          \node[state] (j2) at (1,0) {*};
          \node[state] (b2) at (2,-1) {b};
        \end{scope}
        \node[state] (i3) at (4,0) {*};
        \foreach \n in {1,2} {
          \pgfmathsetmacro\nPlusOne{int(\n+1)}
          \path[transition] (i\n) to (a\n);
          \path[transition] (i\n) to (j\n);
          \path[transition] (j\n) to (b\n);
          \path[transition] (j\n) to (i\nPlusOne);
        }
        \node (dots) at ([xshift=8mm]i3.east) {\cdots};
        \path[transition,-] (i3) to (dots);
      \end{scope}

      \foreach \fitAround [count=\idx] in {{(i1) ([xshift=-2mm]i1.west)}, (j1), (i2), (j2), (i3)} {
        \node[fit={\fitAround},inner sep=0pt,outer sep=0pt] (bbox C\idx) {};
        \pgfmathsetmacro\idxZero{int(\idx-1)}
        \node[coalgname,anchor=south] (C\idx) at (bbox C\idx.north) {\ensuremath{T_\idxZero}};
      }
      \node[coalgname,anchor=base] (C label) at (C1.base -| i.north) {\ensuremath{C}};

      \begin{scope}[on background layer]
      \foreach \fitAround [count=\idx] in {{(i1) ([xshift=-2mm]i1.west)}, (j1), (i2), (j2), (i3)} {
        \node[coalgebra frame,fit={(bbox C\idx) (C\idx) (a1.south -| bbox C\idx.south)},xshift=-2pt]
          (last frame) {};
      }
      \node[coalgebra frame,fit={(scope1) (C label) (scope1.south |- i1.north) (scope1.south |- a1.south)}]
        (cyclic frame)
        {};
      \end{scope}

      \begin{scope}[commutative diagrams/.cd,every diagram]
        \path[commutative diagrams/every arrow,shorten >= 2mm] (dots.east |- i)
          to node[commutative diagrams/every label] {\ensuremath{h}} (cyclic frame.west |- i);
      \end{scope}
    \end{tikzpicture}
    \caption{Tree unravelling $h$ of an $F_\Sigma$-coalgebra for $\Sigma= \set{a/0,b/0,\mathord{*}/2}$.}
    \label{figSigmaUnravel}
  \end{figure}
\end{exa}

\section{Conclusions and Future Work}\label{sec:fw}
Having characterized what it means for a coalgebra to be a tree, it remains
open how other graph-based notions can be described by universal properties.
Here, we think of notions like directed acyclic graphs (DAGs), shortest paths,
and maximum flow.
\twnote{}%

What also remains for future work is to define a coalgebraic recursion
principle on trees that works for possibly infinite trees. Established
coalgebraic recursion principles known from well-founded coalgebras and
recursive coalgebras restrict to finite-depth structures due to their
\textqt{bottom up} strategy~\cite{osius1974,Taylor96,Taylor23,taylor1999,JeanninKS17,AdamekMM20}.

It would also be interesting to connect our coalgebraic tree notion to
trees considered in the open-map approach to bisimilarity, where a tree is
\emph{a colimit of paths}~\cite{DubutGG16}, covering non-deterministic systems and weighted systems~\cite{DubutW23}.
\bknote{}%

Natural transformations $\alpha\colon F\to G$ admit a canonical extension to functors of the type $\Coalg_I(F)\to \Coalg_I(G)$.
It is an open question which sufficient (or even necessary) conditions on the
natural transformation $\alpha$ ensure that its extension preserves the property
of being a tree. This may also give insights into how to relate the coalgebraic notion of trees
in $\Set$ to properties of the canonical graph.
\twnote{}%
\twnote[inline]{}%

\bibliographystyle{alphaurl}
\bibliography{refs}

\end{document}